\documentclass[12pt]{iopart}
\usepackage{epsfig}
\usepackage{graphicx}
\usepackage{amssymb}
\usepackage{bbm}
\usepackage{moreverb}
\usepackage[dvips]{color}
\usepackage[normalem]{ulem}

\usepackage[left=1.54cm,top=1.54cm,bottom=1.54cm,right=1.54cm]{geometry}

\begin{document}

\title[OOE dynamics with MPS]{\begin{center}Out of equilibrium dynamics with Matrix Product States\end{center}}
\author{M. L. Wall$^1$ and Lincoln D. Carr$^{1,2}$}
\address{$^1$Department of Physics, Colorado School of Mines, Golden, CO 80401, USA}
\address{$^2$Universit\"at Heidelberg, Physikalisches Institut, D-69120 Heidelberg, Germany}

\begin{abstract}
Theoretical understanding of strongly correlated systems in one spatial dimension (1D) has been greatly advanced by the density-matrix renormalization group (DMRG) algorithm, which is a variational approach using a class of entanglement-restricted states called Matrix Product States (MPSs).  However, DRMG suffers from inherent accuracy restrictions when multiple states are involved due to multi-state targeting and also the approximate representation of the Hamiltonian and other operators.  By formulating the variational approach of DMRG explicitly for MPSs one can avoid errors inherent in the multi-state targeting approach.  Furthermore, by using the Matrix Product Operator (MPO) formalism, one can exactly represent the Hamiltonian and other operators relevant for the calculation.  The MPO approach allows 1D Hamiltonians to be templated using a small set of finite state automaton rules without reference to the particular microscopic degrees of freedom.  We present two algorithms which take advantage of these properties: eMPS to find excited states of 1D Hamiltonians and tMPS for the time evolution of a generic time-dependent 1D Hamiltonian.  We properly account for time-ordering of the propagator such that the error does not depend on the rate of change of the Hamiltonian.  Our algorithms use only the MPO form of the Hamiltonian, and so are applicable to microscopic degrees of freedom of any variety, and do not require Hamiltonian-specialized implementation.  We benchmark our algorithms with a case study of the Ising model, where the critical point is located using entanglement measures.  We then study the dynamics of this model under a time-dependent quench of the transverse field through the critical point.  Finally, we present studies of a dipolar, or long-range Ising model, again using entanglement measures to find the critical point and study the dynamics of a time-dependent quench through the critical point.
\end{abstract}

\section{Introduction}
\label{sec:introduction}

The great success of experimental ultracold atomic physics has made the study of strongly correlated one-dimensional (1D) quantum systems a major avenue of current physics research.  Examples of novel condensed matter physics realized with 1D atomic systems include the role of integrability in thermalization~\cite{Wenger_Kinoshita_06} and static~\cite{Trotzky_Pollet_10} and dynamic~\cite{Trotzky_Chen_11} quantum simulators of Hubbard models.  Furthermore, as ultracold molecules approach quantum degeneracy~\cite{Ni_Ospelkaus_08, Ospelkaus_Ni_09, Deiglmayr_Grochola_08}, lattice models with complex internal degrees of freedom and long-range interactions become relevant~\cite{Wall_Carr_10,Gorshkov_Manmana_11,Gorshkov_Manmana_11b}.  As more and more complex models become amenable to study, the need for numerical methods which can adapt to different degrees of freedom, different Hamiltonians, and different dynamical processes thus becomes essential.

In addition to practical interest in understanding and benchmarking atomic and molecular quantum simulators, the ability to simulate the dynamics of 1D systems also provides insight into fundamental questions such as the universality of dynamics approach quantum critical points and the effects of integrability on the thermalization process~\cite{Polkovnikov_Sengupta_11}.  The natural setting for studying dynamics near critical points is a quantum quench where one of the parameters of the Hamiltonian is driven through a quantum critical point following a time dependent protocol, for example
\begin{eqnarray}
g\left(t\right)&=&g_0+\frac{v\left(t-t_0\right)^r}{r!}\theta\left(t-t_0\right)\, ,
\end{eqnarray}
with $\theta\left(t\right)$ the step function.  Such quenches pose a special difficulty for numerical studies as by definition they involve evolution with a time-dependent Hamiltonian which does not commute with itself at different times.  The propagator is then generally a time-ordered exponential whose precise form may be difficult to ascertain.  Standard methods such as the Suzuki-Trotter expansion which ignore the time dependence of the Hamiltonian require~\cite{Poulin_Qarry_11} that the infinitesimal time step used be much less than the fluctuation time-scale of $H\left(t\right)$ to be valid, $\delta t\ll \left|\partial H/\partial t\right|^{-1}$.  This can cause simulations with rapid quench rates to become numerically very costly, and invalidates the approach altogether for non-analytic time dependence.

Currently the only unbiased method available for the dynamics of quantum systems is exact diagonalization (ED).  By unbiased, we refer to the fact that the other methods available for dynamics are generally variational, and so have a bias towards a particular ansatz.  ED is limited in an essential way by the exponential growth of the size of the Hilbert space with the physical size of the system.  The current state of the art is $\sim 40$ spins for spin-$1/2$ models and $20$ sites for a fermionic Hubbard model at half filling.  These sizes are often too small for accurate finite-size scaling.  An extremely powerful method for the low-energy properties of 1D systems is White's Density Matrix Renormalization Group (DMRG) algorithm, which uses a variational ansatz based on a class of states known as Matrix Product States (MPSs).  MPSs will be reviewed in Sec.~\ref{sec:MPS}.  DMRG uses an iterative procedure to develop a set of reduced bases that the full many-body problem is projected into, and then variationally minimizes an energy functional in this reduced space, enlarging it if necessary.  DMRG uses an implicit MPS representation, which is to say that the state is not stored explicitly.  This also means that the Hamiltonian and other operators in the calculation are stored in an approximate way, as they are represented within the reduced basis describing the variational state.  This does not cause problems in practice when a single state is sought using the DMRG process.  In fact, one can show that the algorithm to variationally find the ground state is identical when phrased in the implicit formulation of DMRG and when using an explicit MPS representation for the variational state, other than the representation of the Hamiltonian~\cite{Schollwoeck_11}.  However, because of the exact representation of operators independent of the state, MPSs can put rigorous bounds on distances from exact quantum states such as eigenstates by considering quantities like the variance $\langle\psi| \left(\hat{H}-E\right)^2|\psi\rangle$ with $\hat{H}$ the Hamiltonian operator and $E$ the energy expectation of the MPS $|\psi\rangle$.  In contrast, DMRG can only return the distance of the variational state from the approximate operator $\hat{H}$ in the given variational basis, and is unable to determine how well the given variational basis represents the true operator.  A particularly clear indication of the failures this can cause is given in Ref.~\cite{Garcia-Ripoll_06} where time evolution of a particular initial state in DMRG fails because the Hamiltonian has no nonzero matrix elements in the initial DMRG basis.

The situation becomes much different when multiple states are sought using the DMRG procedure.  In this case the reduced density matrix used to determine the optimal reduced bases for the algorithm is a convex sum of the reduced density matrices for the desired states.  This is called \emph{multi-state targeting}.  In contrast, an explicit MPS representation stores each of the desired states separately as an MPS.  In multi-state targeting, none of the states can be represented with the same accuracy available if DMRG targeted that state alone.  The MPS representation also deals automatically with the fact that each state has its own optimal bases for representation, whereas in DMRG these bases are all tied together by the multi-state targeting.  In this work we present two algorithms which take advantage of MPSs' ability to deal with multiple states, eMPS to find excited states of 1D Hamiltonians and tMPS to simulate the dynamics of a generic time-dependent Hamiltonian.  In the first algorithm a projector orthogonal to a set of lower-lying eigenstates is constructed from their MPS representations and used to orthogonalize a variational state against this set.  In the second algorithm Krylov vectors in a Lanczos approximation to the matrix exponential are stored separately as MPSs and combined in an optimal way only at the end of the calculation.  While Krylov-based MPS approaches have been used~\cite{Garcia-Ripoll_06,Keilmann_GarciaRipoll_08} to study time-dependent processes, the errors in these approaches were set by time derivative of the Hamiltonian.  In contrast, by taking explicit account of the time ordering of the propagator, the errors in our approach are set only by commutators of the Hamiltonian at different times, and hence allow for larger time steps.  Because of the explicit MPS representation, we are able to put bounds on the errors in each step of the calculations.

Finally, MPSs have a natural operator-valued extension known as Matrix Product Operators (MPOs) which allow for the exact representation of all operators used in the calculation.  We present a general framework for constructing MPOs from a set of rules which is independent of the nature of the microscopic degrees of freedom.  This allows for the templating of 1D Hamiltonians for general purpose software.  In addition, the ability to perform arithmetic operations on MPOs exactly enables us to perform time-evolution using our tMPS algorithm with knowledge only of the MPO form of the Hamiltonian and the time-dependent functional form of its parameters.  To emphasize the general nature of our algorithms, we include a generic simulation protocol for the out-of-equilibrium dynamics of strongly correlated 1D systems using the algorithms presented in this paper.

The remainder of this paper is organized as follows.  In Sec.~\ref{sec:BFoMPF} we review the theory of MPSs, MPOs, and their canonical forms.  In addition to providing a canonical form for operators within the matrix product formalism, we define finite state automaton rules for MPOs and demonstrate how 1D Hamiltonians can be constructed from a small set of such rules.  In Sec.~\ref{sec:vMPS} we review the algorithm for finding ground states using MPSs as variational ans\"atze, and in Sec.~\ref{sec:eMPS} we present the eMPS algorithm which extends the ground state search to general excited states.  Sec.~\ref{sec:Obs} discusses how to extract observable quantities from MPSs.  In Sec.~\ref{sec:TE} we discuss methods for time evolution with MPS.  In particular, we provide the tMPS algorithm to time evolve an MPS using only the MPO representation of a Hamiltonian and the functional form of its time-dependent parameters.  We contrast our approach with other Krylov subspace approaches and identify the possible sources of error.  In Sec.~\ref{sec:CaseStudies} we present two case studies.  The first is of the Ising model in a transverse field, where we study both the statics and the dynamics of a linear quench of the transverse field through the quantum critical point.  The second is of a dipolar, or long-range Ising model in a transverse field, where we also determine the critical point from the statics and study a linear quench of the transverse field.  Finally, in Sec.~\ref{sec:concl}, we conclude.  Details concerning numerically exact solutions for the Ising model which are used to benchmark our algorithms are given in \ref{sec:IsingExact}.

\section{Brief review of Matrix Product Formalism}
\label{sec:BFoMPF}

\subsection{Matrix Product States}
\label{sec:MPS}

The Hilbert space of a quantum mechanical many-body system is exponentially large in the physical size of the system, for example the number of unit cells in a lattice or the number of particles.  Stated another way, a state picked at random from the Hilbert space of a quantum many-body system will have entanglement (as quantified by the Schmidt measure~\cite{Eisert_Briegel_01}) which grows exponentially with the system size.  In contrast to this random state, it has been shown that the class of states which are physically relevant in the sense that they can be prepared from some reference state by generic time evolution in polynomial time~\cite{Poulin_Qarry_11} or are useful for quantum computation~\cite{Gross_Flammia_09} is much smaller than the full Hilbert space.  In 1D, the physically relevant class of states appears to be those which have entanglement which is either constant or polynomially growing as a function of system size.  A convenient representation of states with entanglement restricted in this manner is known as \emph{matrix product states} (MPSs)~\cite{oestlund_rommer_95, oestlund_rommer_97,Schollwoeck_11}.

We consider our physical system to comprised of a 1D lattice of \emph{sites}, where each site $i$ is a $d$-dimensional Hilbert space $\mathbb{H}_i$ spanned by the vectors $\left\{|i\rangle\right\}$.
  We will refer to $d$ as the \emph{local dimension}, and take all sites to be identical for simplicity.  We define an MPS on a lattice with $L$ sites as
\begin{eqnarray}
\label{eq:MPSdef}|\psi_{\mathrm{MPS}}\rangle=\sum_{i_1,\dots i_L=1}^{d}\mathrm{Tr}\left(A^{\left[1\right]i_1}\dots A^{\left[L\right]i_L}\right)|i_1\dots i_L\rangle\, ,
\end{eqnarray}
where the object $A^{[k]i_k}$ is a $\chi_{k}\times \chi_{k+1}$ matrix (with $\chi_1=\chi_{L+1}$) and $\mathrm{Tr}$ denotes the matrix trace.  We will refer to the maximum linear dimension of any of the matrices $A^{[k]i_k}$, $\max_k \chi_k$, as the \emph{bond dimension} of the MPS, and denote this quantity by $\chi$.  $\chi$ may be used as a measure of the entanglement of the state~\cite{Schollwoeck_11}.  In this work we will focus on systems with open boundary conditions (OBC).  MPS algorithms can also be devised for systems with periodic boundary conditions, as discussed in Refs.~\cite{verstraete_porras_04,pippan_white_10,Wall_Carr_10,Rossini_Giovannetti_11}, but these algorithms have worse scaling and are generally less numerically stable than their OBC counterparts.  For OBC, $\chi_1=\chi_{L+1}=1$, and arguments using the Schmidt decomposition demonstrate that $\chi_k\le \min(d^{k-1},d^{L-k})$~\cite{perezgarcia_verstraete_07}.

MPSs have been used for many years to represent exact ground states of parent Hamiltonians~\cite{Auerbach_94} which are formed from projectors onto local high-symmetry subspaces~\cite{Affleck_Kennedy_87,Tu_Zhang_08}.  However, it was not until the great success of the density matrix renormalization group algorithm (DMRG) pioneered by White~\cite{white_92} that MPSs became valuable as variational \emph{ans\"atze} in their own right~\cite{verstraete_porras_04}.  Why are MPSs useful as variational \emph{ans\"atze}?  It has been shown~\cite{Hastings_07} that the ground states of gapped 1D systems have bipartite entanglement which does not depend on the system size.  Such states can be represented exactly as MPSs with a fixed bond dimension~\cite{verstraete_cirac_06}.  This is an example of an \emph{area law}~\cite{Eisert_Cramer_10}; the entanglement between two disjoint subsystems depends only on the boundary of the two regions and not on their volume.  For systems near a quantum critical point which is described by a conformal field theory (CFT), this area law is subject to weak logarithmic violations such that the entropy of entanglement between two subsystems of size $L$ is given by the Calabrese-Cardy formula~\cite{Holzhey_Larsen_94, Calabrese_Cardy_04}
\begin{eqnarray}
\label{eq:CC}S_L\sim a+\frac{c}{6}\log L\, ,
\end{eqnarray}
where $a$ is a constant and $c$ is the central charge of the underlying CFT.  Here $\sim$ denotes scaling equivalence in the bulk of an infinite system.  In finite systems there are often oscillating boundary and finite size contributions~\cite{Affleck_Laflorencie_09,Calabrese_Cardy_09}.  Hence, the bond dimension of an MPS describing a conformally invariant critical system is given as $\chi_L\sim \exp S_L\sim e^a L^{c/6}$, which grows only polynomially in the system size.  Typical values of $c$ range from 1/2 for the Ising model~\cite{Cardy_96} and 1 for the Bose-Hubbard model~\cite{Laeuchli_Kollath_08} to 2 for more exotic phases like the gapless Mott insulator of the JK model~\cite{Mishmash_Block_11}.  We note that, strictly speaking, finding an MPS which approximates the ground state of an arbitrary 1D Hamiltonian to an accuracy which is an inverse polynomial in the system size is still NP-complete~\cite{Schuch_Cirac_08}, but practical experience demonstrates that this method is extremely useful and robust for physical systems of interest.

We adopt the following conventions for the representation of tensors: we use roman indices for indices which correspond to physical states and greek indices for indices which are summed over in the matrix-product ansatz.  Explicitly writing out Eq.~(\ref{eq:MPSdef}) with these indices, we have
\begin{eqnarray}
|\psi_{\mathrm{MPS}}\rangle=\sum_{\alpha_1\dots \alpha_{L-1}}\sum_{i_1,\dots i_L=1}^{d}A^{\left[1\right]i_1}_{\alpha_1\alpha_2}\dots A^{\left[L\right]i_L}_{\alpha_{L-1}\alpha_1}|i_1\dots i_L\rangle\, .
\end{eqnarray}
A superscript index in square brackets $\left[\right]$ denotes the lattice site that the physical indices of the tensor describe.  A superscript index in curly braces $\left\{\right\}$ denotes association with a particular many-body state.  For example, the MPS tensors at site $j$ of the MPSs $|\phi_k\rangle$, $k=1,\dots ,n$ would be denoted $A_{\alpha\beta}^{\left[j\right]\left\{k\right\}i_j}$.  Finally, indices which appear together in parentheses, e.g.~$\left(\alpha\beta\right)$, represent a composite index which runs over the Cartesian product of the indices in the parentheses.  As an example, if $\alpha=1,\dots, \chi_{\alpha}$ and $\beta=1,\dots \chi_{\beta}$, $\left(\alpha\beta\right)=(\alpha-1)\chi_{\beta}+\beta$.  To lighten the notation, we will leave off indices when they are unnecessary.

\begin{figure}[t]
\begin{center}
\epsfxsize=0.6\textwidth \leavevmode \epsfbox{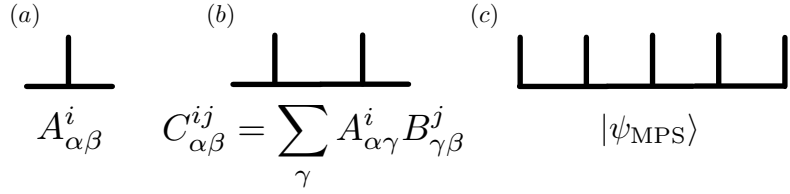}
\caption{  \label{fig:diagrams}  \emph{Diagrammatic notation for tensor networks.} (a) A rank-three tensor is represented as a point with three lines extending from it.  (b) Contraction of two rank-three tensors is accomplished by connecting the contracted index, and produces a tensor of rank four.  (c) An MPS on 5 sites with open boundary conditions }is represented as a contraction over rank three tensors with two rank-two boundary tensors.
\end{center}
\end{figure}

A particularly useful means to visualize MPSs and manipulations with them is provided by \emph{tensor network} diagrams like those shown in Fig.~\ref{fig:diagrams}~\cite{Shi_Duan_06}.  Here a rank-$k$ tensor is represented by a point with $k$ lines extending from it.  Each line represents one of the indices of the tensor.  Whenever a line connects two points, that index is summed over, and disconnected lines represent free indices.  Hence, an MPS can be represented as a chain of rank-$3$ tensors as in Fig.~\ref{fig:diagrams}(c).  Note that the first and last MPS tensors are rank two because we have assumed OBC and so $\chi_1=\chi_{L+1}=1$.

\begin{figure}[t]
\begin{center}
\epsfxsize=0.8\textwidth \leavevmode \epsfbox{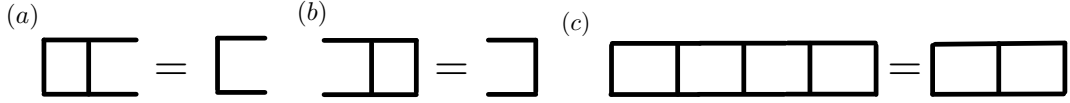}
\caption{  \label{fig:canonical}   \emph{Gauge conditions for MPSs.} (a) Left gauge condition Eq~(\ref{eq:leftcanonical}). (b) Right gauge condition  Eq~(\ref{eq:rightcanonical}). (c) Using (a) and (b), the norm squared of an MPS reduces to the trace of its orthogonality center squared, here chosen to be the third site.}
\end{center}
\end{figure}

We note that the MPS definition Eq.~(\ref{eq:MPSdef}) does not uniquely specify the tensors $A$.  That is, we can insert an invertible matrix $X$ and its inverse $X^{-1}$ between any two MPS matrices without altering the physical content of the state:
\begin{eqnarray}
|\psi_{\mathrm{MPS}}\rangle&=&\sum_{i_1,\dots i_L=1}^{d}\mathrm{Tr}\left(A^{\left[1\right]i_1}\dots A^{\left[L\right]i_L}\right)|i_1\dots i_L\rangle\, ,\\
\tilde{A}^{\left[j\right]i_j}&=& A^{\left[j\right]i_j}X\, ,\;\;\; \tilde{A}^{\left[j+1\right]i_{j+1}}= X^{-1}A^{\left[j+1\right]i_{j+1}}\, ,\\
|\tilde{\psi}_{\mathrm{MPS}}\rangle&=&\sum_{i_1,\dots i_L=1}^{d}\mathrm{Tr}\left(A^{\left[1\right]i_1}\dots \tilde{A}^{\left[j\right]i_j} \tilde{A}^{\left[j+1\right]i_{j+1}}\dots A^{\left[L\right]i_L}\right)|i_1\dots i_L\rangle=|\psi_{\mathrm{MPS}}\rangle\, .
\end{eqnarray}
This is referred to as \emph{gauge freedom} in the literature~\cite{perezgarcia_verstraete_07}.  For OBC, we can specify the state uniquely\footnote{The state is unique up to possible degeneracies in the Schmidt decomposition.} by choosing a site $k$, which we call the \emph{orthogonality center} of the MPS, and requiring that all sites $i$ to the left and right of $k$ satisfy the left
\begin{eqnarray}
\label{eq:leftcanonical}\sum_{i} {A^{i}}^{\dagger}A^i&=I
\end{eqnarray}
and right
\begin{eqnarray}
\label{eq:rightcanonical}\sum_{i} A^i{A^{i}}^{\dagger}&=I
\end{eqnarray}
gauge conditions, respectively.  In these expressions, $I$ is the appropriately dimensioned identity matrix.  These conditions are shown in graphical notation in Fig.~\ref{fig:canonical}(a) and (b), respectively.\footnote{Here we also establish the graphical convention that downwards pointing lines correspond to Hermitian conjugates of tensors.}  Graphically it is clear that the norm squared of the state is
\begin{eqnarray}
\langle \psi|\psi\rangle=\sum_i \mathrm{Tr}\left({A^{\left[k\right]i}}^{\dagger}{A^{\left[k\right]i}}\right)\, ,
\end{eqnarray}
as shown in Fig.~\ref{fig:canonical}(c), and so this site carries all information about the norm of the state.  This particular canonical form for an MPS is called \emph{mixed canonical form}~\cite{Schollwoeck_11}.  The mixed canonical form is crucial for improving the speed and numerical stability of variational algorithms with MPSs.

We can impose the left gauge conditions via the following recursion:
\begin{eqnarray}
\label{eq:l1}\tilde{A}_{\left(\alpha i\right)\beta}&=A^{\left[\ell\right] i}_{\alpha \beta}\, ,\\
\label{eq:lSVD} \sum_{\gamma}U_{\left(\alpha i\right) \gamma}\Sigma_{\gamma} V^{\dagger}_{\gamma\beta}&=\tilde{A}_{\left(\alpha i\right)\beta}\, ,\\
A^{\left[\ell\right] i}_{\alpha \gamma}&=U_{\left(\alpha i\right) \gamma}\, ,\\
\label{eq:lf}A^{\left[\ell+1\right] i}_{\gamma \eta}&=\sum_{\beta}\Sigma_{\gamma} V^{\dagger}_{\gamma\beta}A^{\left[\ell+1\right] i}_{\beta \eta}\, ,
\end{eqnarray}
where Eq.~(\ref{eq:lSVD}) represents the singular value decomposition (SVD) of $\tilde{A}$ with $\Sigma$ a diagonal matrix of singular values and $U$ and $V$ unitary matrices.\footnote{Note that any matrix decomposition of $\tilde{A}$ which returns a unitary matrix as part of the decomposition will suffice in place of the SVD.  In particular, the QR decomposition~\cite{Golub_VanLoan_96} is particularly efficient when the rank of $\tilde{A}$ is not required.}  Because $U$ returned from the SVD is unitary, Eq.~(\ref{eq:leftcanonical}) is satisfied by construction.  Similarly, the recursion for the right gauge conditions is
\begin{eqnarray}
\label{eq:r1}\tilde{A}_{\alpha \left(i \beta\right)}&=A^{\left[\ell\right] i}_{\alpha \beta}\, ,\\
 \sum_{\gamma}U_{\alpha \gamma}\Sigma_{\gamma} V^{\dagger}_{\gamma\left(i \beta\right)}&=\tilde{A}_{\alpha \left(i \beta\right)}\, ,\\
A^{\left[\ell\right] i}_{\gamma\left(i\beta\right)}&=V^{\dagger}_{\gamma\left(i \beta\right)} \, ,\\
\label{eq:rf}A^{\left[\ell-1\right] i}_{\eta \gamma }&=\sum_{\alpha}A^{\left[\ell-1\right] i}_{\eta \alpha }U_{\alpha \gamma}\Sigma_{\gamma}\, .
\end{eqnarray}
To put a general state into mixed canonical form with orthogonality center $k$ we begin at site 1 and iterate Eqs.~(\ref{eq:l1})-(\ref{eq:lf}) until we reach site $k$, then start at site $L$ and iterate Eqs.~(\ref{eq:r1})-(\ref{eq:rf}) until we again reach site $k$.

Finally, we note that the set of all MPSs with a fixed bond dimension $\chi$ is not a vector space, as the sum of two MPSs with bond dimensions $\chi_A$ and $\chi_B$ has a bond dimension $\chi$ which is bounded by the sum of the two bond dimensions $\chi\le \chi_A+\chi_B$.  This can be seen from considering the sum of the two states $|0\dots 0\rangle$ and $|1\dots 1\rangle$, with MPS representations
\begin{eqnarray}
|0\dots 0\rangle&=&\sum_{i_1\dots i_L}A^{i_1}\dots A^{i_L}|i_1\dots i_L\rangle\, ,\;\;\; A^i=\left(\delta_{i,0}\right)\, ,\\
|1\dots 1\rangle&=&\sum_{i_1\dots i_L}B^{i_1}\dots B^{i_L}|i_1\dots i_L\rangle \,, \;\;\; B^{i}=\left(\delta_{i,1}\right)\, .
\end{eqnarray}
The matrices $A^i$ and $B^i$ have a bond dimension of 1, as these are product states.  Their sum is
\begin{eqnarray}
|0\dots 0\rangle+|1\dots 1\rangle=\sum_{i_1\dots i_L}C^{\left[1\right]i_1}C^{\left[2\right]i_2} \dots C^{\left[L-1\right]i_{L-1}}C^{\left[L\right] i_L}|i_1\dots i_L\rangle\, ,\\
C^{\left[1\right]i}=\left(\begin{array}{cc} \delta_{i0}&\delta_{i1}\end{array}\right)\,; \;\;\; C^{\left[j\right]i}=\left(\begin{array}{cc} \delta_{i0}&0\\ 0&\delta_{i1}\end{array}\right)\,, 2\le j\le L-1\, ;\;\;\; C^{\left[L\right]i}=\left(\begin{array}{c} \delta_{i0}\\ \delta_{i1}\end{array}\right)\, ;
\end{eqnarray}
which has a bond dimension of 2.

\subsection{Matrix Product Operators}
\label{sec:MPO}
\begin{figure}[t]
\begin{center}
\epsfxsize=0.6\textwidth\leavevmode \epsfbox{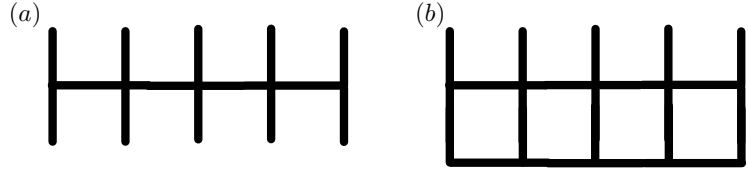}
\caption{  \label{fig:MPOdiagrams}   \emph{MPO in diagrammatic notation.} (a) A matrix product operator consists of a contraction of rank-four tensors.  We adopt the graphical convention that the line below the horizontal corresponds to $i'$ and the line above the horizontal to $i$ for an operator $O_{i i'}$ which takes the local state from $|i'\rangle$ to $|i\rangle$.  (b) The product of an MPO and an MPS produces another MPS whose bond dimension is the product of the bond dimensions of the original MPS and the MPO.}
\end{center}
\end{figure}
The natural operator generalization of MPSs is a Matrix Product Operator (MPO), defined as
\begin{eqnarray}
\label{eq:MPO} \hat{O}_{\mathrm{MPO}}=\sum_{i_1,\dots i_L=1}^{d}\sum_{i_1',\dots i_L'=1}^{d}\mathrm{Tr}\left({W}^{\left[1\right]i_1i_1'}\dots {W}^{\left[L\right]i_Li_L'}\right)|i_1\dots i_L\rangle\langle i_1'\dots i_L'|\, .
\end{eqnarray}
Here $W^{\left[k\right]ii'}$ is a $\chi^O_{k}\times \chi^O_{k+1}$ dimensional matrix, and we will again refer to the maximum value of $\chi^O$ as the bond dimension of the operator.  Note that this bond dimension $\chi^O$ need not be the same as the bond dimension $\chi$ appearing in the MPS representation, Eq.~(\ref{eq:MPSdef}).  That an MPO takes MPSs to MPSs can be seen clearly from the graphical representation of Fig.~\ref{fig:MPOdiagrams}.  We also see from this representation that the bond dimension of the MPS representing the product of an MPO and an MPS is the product of the bond dimensions of the MPO and the MPS.  Because the states $|i_1\dots i_L\rangle$ are tensor products, we can also use the notation
\begin{eqnarray}
\hat{O}_{\mathrm{MPO}}=\sum_{i_1,\dots i_L=1}^{d}\sum_{i_1',\dots i_L'=1}^{d}\mathrm{Tr}\left(\mathcal{W}^{\left[1\right]i_1i_1'}\dots \mathcal{W}^{\left[L\right]i_Li_L'}\right)\, ,
\end{eqnarray}
where each one of the objects $\mathcal{W}^{\left[j\right]i_ji_j'}\equiv W^{\left[j\right]i_ji_j'}|i_j\rangle\langle i_j'|$.  That is to say, we can consider the matrices which appear in the matrix-product ansatz of an MPO to be operator-valued.

It is tempting to look for canonical forms for MPOs just as we did for MPSs, but the relevant norm for MPOs is the Frobenius norm $\langle \hat{O}_1,\hat{O}_2\rangle=\mathrm{Tr}(\hat{O}_1^{\dagger}\hat{O}_2)$ which scales exponentially in the local dimension with the number of lattice sites.  Thus, for a typical many-body system with an exponentially large Hilbert space, the elements of the orthogonality center can differ in magnitude greatly, causing a catastrophic loss of precision during orthogonalization.  However, most physically relevant MPOs such as one-dimensional Hamiltonians can be written down exactly in terms of an MPO canonical form which is analogous to an LU decomposition~\cite{McCulloch_07}.  To discuss this canonical form, it is useful to recast an MPO as a \emph{finite state automaton} (FSA)~\cite{Osborne_06,Crosswhite_Bacon_08}.  

To recast an MPO as a FSA, we first enumerate all of the physical operators we use to define our local Hilbert space, $\mathcal{O}=\{\hat{O}^{\alpha}\}$, where $\alpha$ labels distinct operators.  We will call this set our \emph{operator alphabet} (OA).  As examples, the OA for the Ising model would be $\left\{\hat{I},\hat{\sigma}^x,\hat{\sigma}^z\right\}$ with $\hat{I}$ the identity operator and $\hat{\sigma}^{\nu}$ the Pauli operator along the $\nu^{\mathrm{th}}$ Cartesian direction.  The OA for the Bose-Hubbard model~\cite{fisher1989,jaksch_bruder_98} would be $\left\{\hat{I},\hat{b}^{\dagger},\hat{b},\hat{n}\right\}$, where $\hat{b}$ is a bosonic destruction operator and $\hat{n}=\hat{b}^{\dagger}\hat{b}$ the number operator.  The particular matrix representation of the OA fixes the local basis states $\left\{|i\rangle\right\}$.

Using the OA, we now introduce a set of FSA \emph{rules} $\{\mathcal{R}_p\left(\{\hat{O}^{p_1},\dots,\hat{O}^{p_n}\},\left\{h^p\right\},w^p\right)\}$ which generate strings of the $n$ operators $\{\hat{O}^{p_1},\dots,\hat{O}^{p_n}\}\in\mathcal{O}$ weighted by scalar $w_p$ and variables $\left\{h_p\right\}$.  We will call the variables $\left\{h^p\right\}$ \emph{Hamiltonian parameters} and the scalars $w^p$ \emph{weights}.  Each rule consists of three operator-valued matrices in the case where the  $\left\{h_p\right\}$ do not depend on position.  The first matrix is the rightmost matrix in the MPO representation, $\mathcal{W}^{\left[L\right]}$, and represents the initial configuration.  Next, we have the MPO matrix of the bulk of the chain $\mathcal{W}^{\left[j\right]}$, $2\le j\le L-1$, which takes an input vector of operators on $k$ sites and produces a vector of operators on $k+1$ sites according to some deterministic pattern.  Finally, we have the leftmost matrix in the MPO representation, $\mathcal{W}^{\left[1\right]}$, which extends the operators according to the patten of $\mathcal{W}^{\left[j\right]}$ and returns a $1\times 1$ operator-valued matrix.  The trace of this matrix as in Eq~(\ref{eq:MPO}) is the desired Hamiltonian term.  The generalization to position-dependent Hamiltonian parameters requires $L-2$ matrices in place of the bulk matrix $\mathcal{W}^{\left[j\right]}$, $2\le j\le L-1$, but the only modification is that $h_p$ becomes $h_p\left(j\right)$.

As a concrete example, consider the Ising model
\begin{eqnarray}
\hat{H}=-J\sum_{\langle i,j\rangle}\hat{\sigma}^z_i\hat{\sigma}^z_j-h\sum_i\hat{\sigma}^x_i\, .
\end{eqnarray}
The Hamiltonian consists of two rules.  The first is a site rule $\mathcal{R}_{\mathrm{site}}(\hat{\sigma}^x,h,-1)$ which generates the string $-h\sum_i\hat{\sigma}_i^{x}$.  The three matrices which provide this rule are
\begin{eqnarray}
\hspace*{-1cm}\mathcal{W}_{\mathrm{site}}^{\left[1\right]}=\left(\begin{array}{cc} -h\hat{\sigma}^x&\hat{I}\end{array}\right)\, ,\;\;\; \mathcal{W}_{\mathrm{site}}^{\left[ 2\le j\le L-1\right]}=\left(\begin{array}{cc} \hat{I}&0\\ -h\hat{\sigma}^x&\hat{I}\end{array}\right)\, ,\;\;\; \mathcal{W}_{\mathrm{site}}^{\left[L\right]}=\left(\begin{array}{c} \hat{I}\\ -h\hat{\sigma}^x\end{array}\right)\, .
\end{eqnarray}
As can be verified, 
\begin{eqnarray}
\prod_{j=L-1}^{k}\mathcal{W}_{\mathrm{site}}^{\left[j\right]}\mathcal{W}_{\mathrm{site}}^{\left[L\right]}&=&\left(\begin{array}{c} \hat{I}\dots \hat{I}\\ -h\sum_{i=k}^{L}\hat{\sigma}_i^x\end{array}\right)\, ,
\end{eqnarray}
and so this rule produces the desired operator.  Similarly, there is a bond rule $\mathcal{R}_{\mathrm{bond}}(\{\hat{\sigma^z},\hat{\sigma}^z\},J,-1)$ given by
\begin{eqnarray}
\hspace*{-1cm}\mathcal{W}_{\mathrm{bond}}^{\left[1\right]}=\left(\begin{array}{ccc} 0&-J\hat{\sigma}^z&\hat{I}\end{array}\right)\, ,\;\;\; \mathcal{W}_{\mathrm{bond}}^{\left[ 2\le j\le L-1\right]}=\left(\begin{array}{ccc} \hat{I}&0&0\\ \hat{\sigma}^z&0&0\\ 0&-J\hat{\sigma}^z&\hat{I}\end{array}\right)\, ,\;\;\; \mathcal{W}_{\mathrm{bond}}^{\left[L\right]}=\left(\begin{array}{c} \hat{I}\\ \hat{\sigma}^z\\ 0\end{array}\right)\, ,
\end{eqnarray}
which produces $-J\sum_{\langle i,j\rangle}\hat{\sigma}_i^z\hat{\sigma}_j^z$, with $\langle i,j\rangle$ denoting a sum over nearest neighbors $i$ and $j$.  The full Hamiltonian is given by the direct sum of the matrices.  Collecting rows of the direct sum which are exactly parallel, we have the MPO representation of the full operator
\begin{eqnarray}
\hspace*{-2.5cm}\mathcal{W}_{\mathrm{Ising}}^{\left[1\right]}=\left(\begin{array}{ccc} -h\hat{\sigma}^x&-J\hat{\sigma}^z&\hat{I}\end{array}\right)\, ,\;\;\; \mathcal{W}_{\mathrm{Ising}}^{\left[ 2\le j\le L-1\right]}=\left(\begin{array}{ccc} \hat{I}&0&0\\ \hat{\sigma}^z&0&0\\ -h\hat{\sigma}^x&-J\hat{\sigma}^z&\hat{I}\end{array}\right)\, ,\;\;\; \mathcal{W}_{\mathrm{Ising}}^{\left[L\right]}=\left(\begin{array}{c} \hat{I}\\ \hat{\sigma}^z\\ -h\hat{\sigma}^x\end{array}\right)\, .
\end{eqnarray}
This construction can be readily extended to general sums of nearest-neighbor and on-site interactions with Hamiltonian
\begin{eqnarray}
\hat{H}=-\sum_{\alpha=1}^{n_b}J_{\alpha}\sum_{\langle i,j\rangle}\hat{O}^{b_{1;\alpha}}_i\hat{O}^{b_{2;\alpha}}_j-\sum_{\beta=1}^{n_s}h_{\beta}\sum_i\hat{O}^{s_{\beta}}_i
\end{eqnarray}
by summing up the individual rules:
\begin{eqnarray}
\hspace*{-2.5cm}\mathcal{W}^{\left[1\right]}=\left(\begin{array}{ccccc} -\sum_{\beta=1}^{n_s} h_{\beta}\hat{O}^{s_{\beta}}&-J_1\hat{O}^{b_{1;1}}&\dots &-J_{n_b}\hat{O}^{b_{1;n_b}}&\hat{I}\end{array}\right)\\
\hspace*{-2.5cm}\mathcal{W}^{\left[ 2\le j\le L-1\right]}=\left(\begin{array}{ccccc} 
\hat{I}&0&\dots&0&0\\  
\hat{O}^{b_{2;1}}&0&\dots&0&0\\ 
\vdots&\vdots&\ddots&\vdots&\vdots\\
\hat{O}^{b_{2;n_b}}&0&\dots&0&0\\
-\sum_{\beta=1}^{n_s} h_{\beta}\hat{O}^{s_{\beta}}&-J_1\hat{O}^{b_{1;1}}&\dots &-J_{n_b}\hat{O}^{b_{1;n_b}}&\hat{I}\\
\end{array}\right)\, ,\;\;\;
 \mathcal{W}^{\left[L\right]}=\left(\begin{array}{c} \hat{I}\\  \hat{O}^{b_{2;1}}\\ \vdots\\  \hat{O}^{b_{2;n_b}} \\  -\sum_{\beta=1}^{n_s} h_{\beta}\hat{O}^{s_{\beta}} \end{array}\right)\, .
\end{eqnarray}
The bond dimension of the MPO representation of this Hamiltonian is $2+n_B$.  MPO representations are not restricted to nearest-neighbor terms.  Exponentially decaying terms of the form\footnote{Note in this form that the nearest neighbor coupling is $J$ and only longer ranged couplings contain $\lambda$.}
\begin{eqnarray}
\hat{H}=-J\sum_{i<j}e^{-\lambda (j-i-1)}\hat{O}^1_i\hat{O}^2_j
\end{eqnarray}
 can also be accommodated with the rule $\mathcal{R}_{\mathrm{exp}}(\{\hat{O}^1,\hat{O}^2\},\{J,\lambda\},-1)$
 \begin{eqnarray}
\hspace*{-1cm}\mathcal{W}_{\mathrm{exp}}^{\left[1\right]}=\left(\begin{array}{ccc} 0&-J\hat{O}^1&\hat{I}\end{array}\right)\, ,\;\;\; \mathcal{W}_{\mathrm{exp}}^{\left[ 2\le j\le L-1\right]}=\left(\begin{array}{ccc} \hat{I}&0&0\\ \hat{O}^2&e^{-\lambda}\hat{I}&0\\ 0&-J\hat{O}^1&\hat{I}\end{array}\right)\, ,\;\;\; \mathcal{W}_{\mathrm{exp}}^{\left[L\right]}=\left(\begin{array}{c} \hat{I}\\ \hat{O}^2\\ 0\end{array}\right)\, .
\end{eqnarray}
We can also produce a general monotonically decaying term
\begin{eqnarray}
\label{eq:purefunction}\hat{H}=J\sum_{i<j}f\left(j-i\right)\hat{O}^1_i\hat{O}^2_j
\end{eqnarray}
by approximating the term as a sum of exponentials to a desired distance $r_{\mathrm{cutoff}}$.  That is, we minimize the functional
\begin{eqnarray}
\label{eq:abfunctional}F\left(\mathbf{a},\mathbf{b}\right)=\sum_{i=1}^{n_{\mathrm{exp}}}\sum_{x=1}^{r_{\mathrm{cutoff}}}\left|f\left(x\right)-a_ib_i^{x-1}\right|^2
\end{eqnarray} 
with respect to the exponential weights $a_i$ and decay parameters $b_i$, where $\mathbf{a}$ and $\mathbf{b}$ are the elements $a_i$ and $b_i$, respectively, arranged as vectors and $n_{\mathrm{exp}}$ is a convergence parameter controlling the number of exponentials used in the expansion~\cite{pirvu_murg_10,Crosswhite_Doherty_08,Froewis_Nebendahl_10}.  Note that the fit is only guaranteed to be accurate to $r_{\mathrm{cutoff}}$ while the term has infinite range.  This does not cause difficulties in practice for decaying functions, as the resulting fit is also guaranteed to be monotonically decaying.  The decaying function rule is then just a sum of these $n_{\mathrm{exp}}$ exponential rules, $\mathcal{R}_{\mathrm{df}}(\{\hat{O}^1,\hat{O}^2\},\{J,f(x),n_{exp},r_{\mathrm{cutoff}}\},1)=\sum_{i=1}^{n_{\mathrm{exp}}}\mathcal{R}_{\mathrm{exp}}(\{\hat{O}^1,\hat{O}^2\},\{Ja_i,-\log b_i\},1)$.  In contrast to the other rules presented above this rule is not exact, but the number of exponentials can be increased to any desired accuracy\footnote{By accuracy we mean that the functional Eq.~(\ref{eq:abfunctional}) is smaller than a given tolerance.}-a few tens of exponentials suffice for a $1/r^3$ interaction on a lattice of 50 sites~\cite{Froewis_Nebendahl_10}.  While the bond dimension increases linearly with the number of exponentials, the number of nonzero terms in the MPO also grows only linearly with the number of exponentials, and so operations with the MPO scale well as $n_{\mathrm{exp}}$ increases.  However, the eigenstate of a Hamiltonian with larger $n_{\mathrm{exp}}$ may be more highly entangled than with smaller $n_{\mathrm{exp}}$, leading to an increase in $\chi$ and longer runtimes.  That is to say, the relationship between the bond dimension of an MPO and the bond dimension of an MPS representing an eigenstate of the MPO is difficult to predict.

While the pure functional interaction Eq.~(\ref{eq:purefunction}) is appealing from a theoretical point of view, in practical applications infinite range interactions do not represent a consistent level of approximation.  That is to say, at some distance interactions are screened, where the screening length is set by energetics or lifetime constraints in the case of ultracold atoms or molecules~\cite{wallforthcoming}.  Hence, we provide the finite-ranged rule $\mathcal{R}_{\mathrm{FiniteFunction}}(\{\hat{O}^1,\hat{O}^2\},\{h,f,r_{\mathrm{cutoff}}\},w)$ which generates the Hamiltonian
\begin{eqnarray}
\hat{H}=wh\sum_{i}\sum_{j=i+1}^{i+r_{\mathrm{cutoff}}}f\left(j-i\right)\hat{O}^1_i\hat{O}^2_j
\end{eqnarray}
and is given by
\begin{eqnarray}
\hspace*{-2.5cm}\mathcal{W}^{\left[1\right]}=\left(\begin{array}{cccccccc} 0&0&\dots&0&\dots&0&whf\left(1\right)\hat{O}^1&\hat{I}\end{array}\right)\\
\hspace*{-2.5cm}\mathcal{W}^{\left[2\le j\le L-1\right]}=\left(\begin{array}{cccccccc}
 \hat{I}&0&\dots&0&\dots&0&0&0\\
\hat{O}^{2} &0&\dots&0&\dots&0&0&0\\
\hat{O}^{2} &\frac{f\left(r_{\mathrm{cutoff}}\right)}{f\left(r_{\mathrm{cutoff}}-1\right)}\hat{I}&\dots&0&\dots&0&0&0\\
\vdots&\vdots&\ddots&\vdots&\ddots&\vdots&\vdots&\vdots\\
\hat{O}^{2} &0&\dots&\frac{f\left(k\right)}{f\left(k-1\right)}&\dots&0&0&0\\
\vdots&\vdots&\ddots&\vdots&\ddots&\vdots&\vdots&\vdots\\
\hat{O}^{2} &0&\dots&0&\dots&\frac{f\left(2\right)}{f\left(1\right)}\hat{I}&0&0\\
 0&0&\dots&0&\dots&0&whf\left(1\right)\hat{O}^1&\hat{I}
\end{array}\right)\, ,\;\;\;
\mathcal{W}^{\left[L\right]}=\left(\begin{array}{c} \hat{I}\\ \hat{O}^2\\ \hat{O}^{2} \\\vdots\\ \hat{O}^{2} \\ \vdots\\ \hat{O}^{2} \\ 0\end{array}\right)\, .
\end{eqnarray}
For a vector $f$ which contains zero elements a small modification must be made, but the bond dimension remains the same.  

A small set of rules like those presented above allow us to template operators for MPS simulations.  That is, given a method to produce a Hamiltonian from a set of basic rules, we can generate a wide variety of different Hamiltonians by specifying different OAs, weights, and Hamiltonian parameters as inputs to a program.  The resulting MPOs are lower triangular and usually very sparse, and so sparse matrix structures can be used for efficiency.  Beyond the flexibility that MPOs provide to MPS algorithms, they also allow for arithmetic operations such as addition and multiplication to be performed exactly, albeit at the expense of a growing bond dimension.  The sum of two operators expressed as MPOs has MPO matrices which are the direct sums of the constituent matrices, and the product of two MPOs has MPO matrices which are the direct product of the constituent MPO matrices.  Both of these operations preserve the sparse lower triangular structure of the MPOs.  
\section{Variational Ground State Search}
\label{sec:vMPS}
\begin{figure}[t]
\begin{center}
\epsfxsize=0.8\textwidth \leavevmode \epsfbox{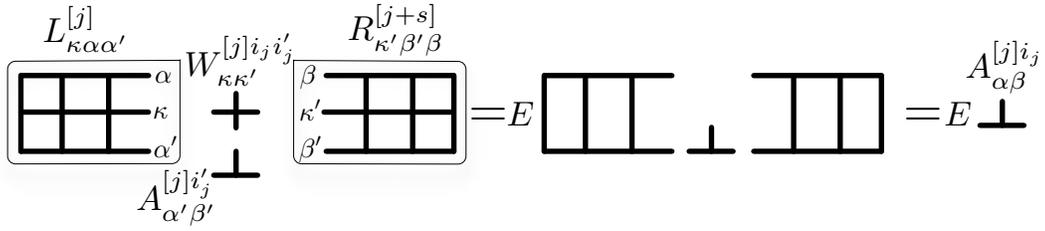}
\caption{  \label{fig:GSdiagrams}  \emph{Variational ground state search in diagrammatic notation.} Here we display the single site ($s=1$) effective Hamiltonian eigenvalue problem for simplicity.  The contractions between the block of tensors to be optimized, the MPO, and the rest of the diagram have been omitted to accentuate the structure of the effective Hamiltonian.  The rightmost equality follows from assuming that the variational site is the orthogonality center.}
\end{center}
\end{figure}

We now turn to using MPSs as variational ans\"atze for the eigenstates of a Hamiltonian expressed as an MPO.  The ground state is found by minimizing the functional
\begin{eqnarray}
\mathcal{E}\left[|\psi\rangle\right]=\langle \psi|\hat{H}|\psi\rangle-E\langle \psi|\psi\rangle
\end{eqnarray}
with respect to the parameters of $|\psi\rangle$, where $E$ is a Lagrange multiplier enforcing normalization.  The general minimization of this functional is an NP-hard problem, so we instead adopt a local search heuristic that has proven to work well in practice.  Let us consider fixing all parameters in the MPS except for a contiguous block of $s$ MPS tensors $A^{\left[j\right]}\dots A^{\left[j+s-1\right]}$.  The single-site ($s=1$) and two-site ($s=2$) algorithms are the most commonly used variants.  We then find the extremum as
\begin{eqnarray}
\frac{\partial}{ \partial {A^{\left[j\right]}}^{\star} \dots {A^{\left[j+s-1\right]}}^{\star}} \left(\langle \psi|\hat{H}|\psi\rangle-E \langle \psi|\psi\rangle\right)=0
\end{eqnarray}
which corresponds to the diagrammatic equation shown in Fig.~\ref{fig:GSdiagrams}.  Here, the partial derivative with respect to a tensor is defined to be a tensor whose elements are the partial derivatives with respect to the elements of the tensor.  If we assume that the block of tensors to be optimized contains the orthogonality center then the rightmost diagram reduces to the block of tensors being optimized and the leftmost diagram is the action of the \emph{effective Hamiltonian} on this same block.  Thus, minimization consists of finding the eigenvector corresponding to the smallest eigenvalue of the effective Hamiltonian eigenvalue problem
\begin{eqnarray}
\label{eq:EHEP}\hat{H}_{\mathrm{eff}}^{\left[j\right]} A^{\left[j\right]}\dots A^{\left[j+s-1\right]}=E A^{\left[j\right]}\dots A^{\left[j+s-1\right]}\, ,
\end{eqnarray}
where the effective Hamiltonian is
\begin{eqnarray}
\label{eq:EH} \hat{H}_{\mathrm{eff}\;\alpha \beta\alpha'\beta'}^{\left[j\right]i_j\dots i_{j+s-1} i_j'\dots i_{j+s-1}'}=\sum_{\kappa\kappa'}\sum_{\gamma_1\dots \gamma_s} L^{\left[j\right]}_{\kappa \alpha \alpha'}W^{\left[j\right]i_ji_j'}_{\kappa\gamma_1} \dots W^{\left[j+s-1\right]i_{j+s-1} i_{j+s-1}'}_{\gamma_s\kappa'} R^{\left[j+s\right]}_{\kappa'\beta'\beta}\, ,
\end{eqnarray}
 $L$ and $R$ are the partial overlaps of the Hamiltonian MPO with the state as in Fig.~\ref{fig:GSdiagrams}, and $W^{\left[j\right]}$ is the MPO tensor at site $j$ of the Hamiltonian.  $\hat{H}_{\mathrm{eff}}^{\left[j\right]}$ represents the Hamiltonian for the variational degrees of freedom in the block to be optimized with the rest of the state held fixed.  This justifies our use of $E$ as the eigenvalue, as $E$ obtained from the solution of this equation is the current best estimate for the energy.  We can view Eq.~(\ref{eq:EHEP}) as a linear eigenvalue problem by combining indices as
 \begin{eqnarray}
\nonumber \hspace*{-2.0cm}  \sum_{\alpha'i_j'\dots i_{j+s-1}'\beta'}\hat{H}_{\mathrm{eff}\; \left(\alpha i_j\dots i_{j+s-1}\beta\right)\left(\alpha' i_j'\dots i_{j+s-1}'\beta'\right)}\left[A^{\left[j\right]}\dots A^{\left[j+s-1\right]}\right]_{\left(\alpha' i_j'\dots i_{j+s-1}'\beta'\right)}\\
	=E \left[A^{\left[j\right]}\dots A^{\left[j+s-1\right]}\right]_{\left(\alpha i_j\dots i_{j+s-1}\beta\right)}\, .
 \end{eqnarray}
 The linear dimension of this matrix representation of $\hat{H}_{\mathrm{eff}}^{\left[j\right]}$ is $\chi_j d^{s}\chi_{j+s-1} $, and so a solution of this problem with dense methods would require $\mathcal{O}(\chi_j^3 d^{3s}\chi_{j+s-1}^3)$ basic operations, leading to a very slow algorithm of order $\mathcal{O}(\chi^6)$.  In contrast, by taking advantage of the separable form of the effective Hamiltonian Eq.~(\ref{eq:EH}) multiplication of our block of tensors by the effective Hamiltonian can be done in $\mathcal{O}(\chi^3)$ time~\cite{Schollwoeck_11}.  Thus, sparse eigensolvers such as the Lanczos~\cite{Golub_VanLoan_96} or Davidson~\cite{Davidson_75} algorithms, which require only matrix-vector multiplies, should be employed to solve this eigenvalue problem.  

The general algorithm for ground state search is thus as follows.  We begin with an initial state with orthogonality center at site $k$.  We choose a block of tensors containing $k$ and optimize them by solving the effective Hamiltonian eigenvalue problem.  We then shift the orthogonality center and the block of tensors one site to the right and again optimize.  We continue shifting to the right until we reach the right boundary.  We then reverse direction, shifting the orthogonality center and the block of tensors to be optimized  to the left and solving the effective Hamiltonian eigenvalue problem until we reach the left boundary, at which point we reverse direction again.  A single iteration of this optimization cycle which affects each tensor twice is called a \emph{sweep}, and sweeping is continued until convergence.  In addition, using the MPO form of the Hamiltonian, it is possible to develop a caching algorithm for the overlaps $L$ and $R$ such that the solution of this problem requires $\mathcal{O}\left(L\right)$ scaling in the number of lattice sites~\cite{Crosswhite_Bacon_08}.\footnote{This scaling does not account for possible $L$ dependence of the bond dimension $\chi$ such as exists for conformal critical points.}  To do so, we begin the iteration with a guess for the ground state $|\psi\rangle$ assumed to have orthogonality center $k$.  We then use the left recursion
\begin{eqnarray}
L^{\left[1\right]}_{\kappa \alpha\alpha'}&=\delta_{\alpha,1}\delta_{\alpha',1}\delta_{\kappa,1}\, ,\\
\label{eq:LLeftrecur}L^{\left[\ell+1 \right]}_{\kappa\alpha\alpha'}&=\left[\sum_{\gamma' i'}\left[\sum_{i \kappa'}\left[\sum_{\gamma}{A^{\left[\ell\right]i}_{\gamma\alpha}}^{\star}L^{\left[\ell\right]}_{\kappa' \gamma\gamma'}\right]W^{\left[ \ell\right] ii'}_{\kappa'\kappa}\right]A^{\left[\ell\right]i'}_{\gamma'\alpha'}\right]\, ,
\end{eqnarray}
to generate the $L$ overlaps up to $k$ and the right recursion
\begin{eqnarray}
R^{\left[L+1\right]}_{\kappa \beta'\beta}&=\delta_{\beta,1}\delta_{\beta',1}\delta_{\kappa,1}\, ,\\
\label{eq:RRightrecur}R^{\left[\ell \right]}_{\kappa\beta'\beta}&=\left[\sum_{\gamma'i'}A^{\left[\ell\right]i'}_{\beta'\gamma'}\left[\sum_{\kappa'i}W_{\kappa\kappa'}^{\left[\ell\right]ii'}\left[\sum_{\gamma}R_{\kappa'\gamma'\gamma}^{\left[\ell+1\right]}{A^{\left[\ell\right]i}_{ \beta\gamma}}^{\star}\right]\right]\right]\, ,
\end{eqnarray}
to generate the $R$ overlaps down to $k+s$.  Here the square braces indicate the order in which the contraction should be performed to achieve ideal scaling.  Once the eigenvalue problem has been solved and the orthogonality center shifted, we use the recurrence Eq.~(\ref{eq:LLeftrecur}) to update the overlaps when we are sweeping to the right and the recurrence Eq.~(\ref{eq:RRightrecur}) to update the overlaps when we are sweeping to the left.

Convergence is achieved when the variance
\begin{eqnarray}
{\Delta}\equiv \langle \left(\hat{H^2}-E^2\right)\rangle\, ,
\end{eqnarray}
with $E$ the energy eigenvalue, drops below a user-specified tolerance $\epsilon$.  Given the MPO form of the Hamiltonian, the variance operator $\hat{\Delta}\equiv \hat{H^2}-E^2$ can be constructed by constructing an MPO whose matrices $\tilde{W}^{\left[i\right]}$ consist of the direct product of the corresponding matrices from $\hat{H}$, $\tilde{W}^{\left[i\right]}=W^{\left[i\right]}\otimes W^{\left[i\right]}$, and then subtracting $-\hat{I}E^2/L$ from the lower leftmost element of each $\tilde{W}^{\left[i\right]}$, where $\hat{I}$ is the identity operator.  This representation is exact, in contrast to DMRG-based approaches where the basis of the Hamiltonian is tied together with the basis of the state.  The variance is a much better measure of convergence of the \emph{state} than the so-called discarded weight which is used to measure convergence of the two-site DMRG algorithm.  This is because it is a property of the actual MPS state and not of the eigenvalue.  As a note of caution, the variance only guarantees that the state found is \emph{an} eigenstate to the given tolerance, it does not specify that it is the ground state.  This has not proven to cause problems in practice for non-disordered systems.  

In summary, the complete algorithm for variational ground state search is:
\begin{enumerate}
\item \emph{Input}: Input a Hamiltonian $\hat{H}$ in MPO form, an initial guess $|\psi\rangle$ for the ground state in MPS form, and a tolerance $\epsilon$ for the variance. 
\item \emph{Initialization}: Construct the $LR$ overlaps using the recursions Eq.~(\ref{eq:LLeftrecur}) and (\ref{eq:RRightrecur}).
\item \emph{Sweeping} :
\begin{enumerate}
\item Solve the effective Hamiltonian eigenvalue problem Eq.~(\ref{eq:EHEP}) and replace the variational block of tensors with the eigenvector corresponding to the lowest eigenvalue.
\item Shift the orthogonality center to the right and update the $LR$ overlaps using the recursion  Eq.~(\ref{eq:LLeftrecur}).
\item Continue to iterate a and b (\emph{right sweeping}) until the right boundary is reached.
\item Solve the effective Hamiltonian eigenvalue problem Eq.~(\ref{eq:EHEP}) and replace the variational block of tensors with the eigenvector corresponding to the lowest eigenvalue.
\item Shift the orthogonality center to the left and update the $LR$ overlaps using the recursion  Eq.~(\ref{eq:RRightrecur}).
\item Continue to iterate d and e (\emph{left sweeping}) until the left boundary is reached.
\item Iterate a and b until $k$ is reached.
\end{enumerate}
\item \emph{Check convergence}: Using the most recent estimate of the energy eigenvalue $\tilde{E}$ from the last effective Hamiltonian solution, construct the variance operator $\hat{\Delta}$ and find the variance.  If ${\Delta}<\epsilon$, exit, otherwise return to a.
\end{enumerate}

\section{Variational Excited State Search}
\label{sec:eMPS}

\begin{figure}[t]
\begin{center}
\epsfxsize=0.75\textwidth \leavevmode \epsfbox{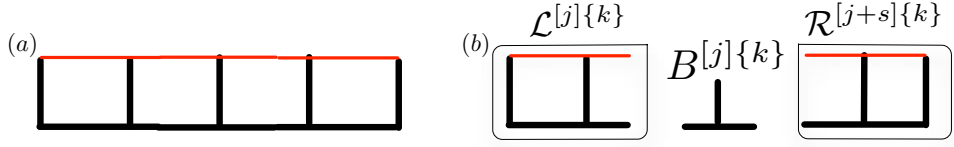}
\caption{  \label{fig:ESdiagrams}  \emph{Linear forms enforcing orthogonality.} (a) The overlap $\langle \psi|\phi_k\rangle$ in diagrammatic notation.  The thin red lines correspond to $\langle \psi|$ and the thick black lines to $|\phi_k\rangle$.  (b) The linear form $F^{\left[j\right]\left\{k\right\}}$ in diagrammatic notation for the single-site case.  As before, we leave the bottom tensor uncontracted to accentuate the definitions of the $\mathcal{LR}$ overlaps.}
\end{center}
\end{figure}

We now turn to finding excited states.  We find the $n^{\mathrm{th}}$ excited state by minimizing the functional
\begin{eqnarray}
\mathcal{E}\left[|\psi\rangle\right]=\langle \psi|\hat{H}|\psi\rangle-E\langle \psi|\psi\rangle-\sum_{k=0}^{n-1}\mu_k\langle \psi |\phi_k\rangle
\end{eqnarray}
where $\{|\phi_k\rangle\}$ are the $n-1$ lower-lying eigenstates of $\hat{H}$ and the $\left\{\mu_k\right\}$ are Lagrange multipliers enforcing the orthogonality constraints $\langle \psi |\phi_k\rangle=0$.  Again fixing a block of $s$ tensors, the minimization of this functional with respect to this block is given by the \emph{projected} effective Hamiltonian eigenvalue problem
\begin{eqnarray}
{\hat{P}^{\left[j\right]}}\,^{\dagger}\hat{H}_{\mathrm{eff}}^{\left[j\right]}{\hat{P}^{\left[j\right]}}A^{\left[j\right]}\dots A^{\left[j+s-1\right]}=E A^{\left[j\right]}\dots A^{\left[j+s-1\right]}\, ,
\end{eqnarray}
where $\hat{P}^{\left[j\right]}$ is a projector into the space orthogonal to the $\{|\phi_k\rangle\}$.  Given the states $\{|\phi_k\rangle\}$ as MPSs, we construct these projectors as follows.  The diagram corresponding to the overlap $\langle \psi|\phi_k\rangle$ is shown in Fig.~\ref{fig:ESdiagrams}(a), with the bold lines corresponding to $|\phi_k\rangle$ and the thin lines to $\langle \psi|$.  This is a \emph{linear} form in all of the MPS tensors of $\langle \psi|$, and so the condition that $|\psi\rangle$ be orthogonal to this state for the given block of tensors with all others held fixed may thus be stated as
\begin{eqnarray}
\label{eq:orthlinearform}\langle \psi|\phi_k\rangle=0&\Rightarrow  \left(\frac{\partial}{ \partial {A^{\left[j\right]}}^{\star} \dots {A^{\left[j+s-1\right]}}^{\star}}\langle \psi|\phi_k\rangle \right)A^{\left[j\right]}\dots A^{\left[j+s-1\right]}=0\\
&\equiv F^{\left[j\right]\left\{k\right\}}A^{\left[j\right]}\dots A^{\left[j+s-1\right]}=0\, .
\end{eqnarray}
Here the linear form enforcing orthogonality $F^{\left[j\right]\left\{k\right\}}$ is shown diagrammatically in Fig.~\ref{fig:ESdiagrams}(b).  We can construct the projector $\hat{P}^{\left[j\right]}$ as
\begin{eqnarray}
\label{eq:prj}\hat{P}^{\left[j\right]}&=\hat{1}-\sum_{kk'}F^{\left[j\right]\left\{k\right\}}\left(\mathcal{N}^{-1}\right)_{kk'}F^{\left[j\right]\left\{k'\right\}}\,^{\dagger}\, ,
\end{eqnarray}
where $(\mathcal{N}^{-1})_{kk'}$ is the $kk'^{\mathrm{th}}$ element of the inverse of the Gram matrix
\begin{eqnarray}
\mathcal{N}_{kk'}=\mathrm{Tr}\left(F^{\left[j\right]\left\{k\right\}}\,^{\dagger} F^{\left[j\right]\left\{k'\right\}}\right)\,.
\end{eqnarray}
This Gram matrix inverse is important to ensure idempotency of the projector.  As before, direct construction of the projected effective Hamiltonian leads to an algorithm which scales poorly as $\mathcal{O}(\chi^6)$.  Hence, it is important to use sparse methods which require only the application of $\hat{P}$ and $\hat{H}$ on some block of tensors $A^{\left[j\right]} \dots {A^{\left[j+s-1\right]}}$.  Direct application of $\hat{P}^{\left[j\right]}$ as written requires $\mathcal{O}(\chi^4)$ operations and also scales quadratically in the number of eigenstates desired $N_E$ due to the double sum in Eq.~(\ref{eq:prj}).  To find a total of $N_E$ eigenstates by this method thus requires $\mathcal{O}(N_E^3\chi^4)$ operations, which is unacceptably slow.  A simple idea to reduce this scaling would be to find the eigenvectors of the inverse Gram matrix and re-express the projectors $F^{\left[j\right]\left\{k\right\}}$ in terms of them, rendering the double sum a single sum.  However, while the Gram matrix $\mathcal{N}$ is Hermitian and positive semidefinite it may also be badly conditioned and singular.  A numerically stable alternative to this idea is to construct the Moore-Penrose pseudoinverse~\cite{Golub_VanLoan_96} of the Gram matrix
\begin{eqnarray}
\left(\mathcal{N}^+\right)_{kk'}&=\sum_{\mu=1}^{n_p}V_{k\mu}\frac{1}{\lambda_{\mu}}V_{k'\mu}^{\star}
\end{eqnarray}
where $V$ is the matrix with the eigenvectors of $\mathcal{N}$ as columns and $\lambda$ are the $n_p$ eigenvalues of $\mathcal{N}$ which are greater than $n\sqrt{\lambda_{\mathrm{max}}}\epsilon$, where $n$ is the linear dimension of $\mathcal{N}$, $\lambda_{\mathrm{max}}$ its largest eigenvalue, and $\epsilon$ the machine precision.  We use this pseudoinverse to transform to a new set of linear forms
\begin{eqnarray}
\label{eq:Gconst}G^{\left[j\right]\left\{\mu\right\}}&=\frac{1}{\sqrt{\lambda_{\mu}}}\sum_kV_{k\mu}F^{\left[j\right]\left\{k\right\}}
\end{eqnarray}
such that
\begin{eqnarray}
\hat{P}^{\left[j\right]}&=\hat{1}-\sum_{\mu}G^{\left[j\right]\left\{\mu\right\}}G^{\left[j\right]\left\{\mu\right\}}\,^{\dagger}\, .
\end{eqnarray}
Often, the dimension of the set $\left\{G^{\left[j\right]\left\{\mu\right\}}\right\}$ is much smaller than $N_E$.  The diagonalization of the Gram matrix requires $\mathcal{O}(N_E^3)$ operations, independent of $\chi$, and its construction and the construction of $G$ in Eq.~(\ref{eq:Gconst}) both require $\mathcal{O}(\chi^2)$ operations.  The operation of $\hat{P}^{\left[j\right]}$ on the variational block of tensors is now
\begin{eqnarray}
\hat{P}^{\left[j\right]}A^{\left[j\right]}\dots A^{\left[j+s-1\right]}&=A^{\left[j\right]}\dots A^{\left[j+s-1\right]}-\sum_{\mu}\mathrm{Tr}\left(G^{\left[j\right]\left\{\mu\right\}}\,^{\dagger}A^{\left[j\right]}\dots A^{\left[j+s-1\right]}\right)G^{\left[j\right]\left\{\mu\right\}}
\end{eqnarray}
which is linear in $N_E$ and scales only as $\mathcal{O}(\chi^2)$.  Thus, the dominant scaling for typical parameters $\chi\gg N_E$ is still the $\mathcal{O}(\chi^3)$ scaling of the effective Hamiltonian multiply, and the algorithm to find $N_E$ excited states scales as $\mathcal{O}(N_E\chi^3)$.  A sweeping approach is used as in the ground state search algorithm, and the iteration is stopped when the variance drops below a user-specified tolerance.  As before, the variance does not guarantee that the state found is the next lowest-lying eigenstate, but this does not usually cause problems in practice.

As with the $LR$ overlaps used in the variational ground state search, one can also cache the overlaps $\mathcal{LR}$ used to construct the linear forms $F$ using the recursions
\begin{eqnarray}
\mathcal{L}^{\left[1\right]\left\{k\right\}}_{\alpha\alpha'}&=\delta_{\alpha,1}\delta_{\alpha',1}\, ,\\
\label{eq:Leftrecur}\mathcal{L}^{\left[\ell+1 \right]\left\{k\right\}}&=\sum_{i}{A^{\left[\ell\right]i}}^{\dagger}\mathcal{L}^{\left[\ell\right]k}B^{\left[\ell\right]\left\{k\right\}i}
\end{eqnarray}
and
\begin{eqnarray}
\mathcal{R}^{\left[L+1\right]k}_{\beta'\beta}&=\delta_{\beta,1}\delta_{\beta',1}\delta_{k,1}\, ,\\
\label{eq:Rightrecur}\mathcal{R}^{\left[\ell \right]\left\{k\right\}}&=\sum_{i}B^{\left[\ell\right]\left\{k\right\}i}\mathcal{R}^{\left[\ell+1\right]\left\{k\right\}}{A^{\left[\ell\right]i}}^{\dagger}
\end{eqnarray}
where $B^{\left[\ell\right]\left\{k\right\}}$ is the MPS tensor of $|\phi_k\rangle$ at site $\ell$.  The linear forms are constructed using these overlaps as
\begin{eqnarray}
\label{eq:Fdef}F^{\left[j\right]\left\{k\right\}i_j\dots i_{j+s-1}}_{\alpha\beta}&=\sum_{\gamma_1\dots \gamma_{s+1}} \mathcal{L}^{\left[j\right] \left\{k\right\}i_j}_{\alpha\gamma_1}B^{\left[j\right] \left\{k\right\}i_j}_{\gamma_1\gamma_2}\dots B^{\left[i+s-1\right] \left\{k\right\}i_{j+s-1}}_{\gamma_s\gamma_{s+1}} \mathcal{R}^{\left[j+s\right]\left\{k\right\}}_{\gamma_{s+1}\beta}\, ,
\end{eqnarray}
see Fig.~\ref{fig:ESdiagrams}(b).

The variational ground state algorithm presented above is essentially equivalent to standard DMRG, aside from the calculation of the variance~\cite{Schollwoeck_11}.  When finding excited states, however, DMRG-based approaches target multiple excited states in a single MPS, which causes the bond dimensions to grow and the quality of each individual eigenstate to degrade.  Furthermore, as the ground state and all excited states are solved together in that approach, the sparse eigensolver must be able to converge interior eigenvalues, which is known to be troublesome~\cite{Golub_VanLoan_96, McCulloch_07}.  We call the present algorithm, which is a sparse and numerically stable variant of that proposed in Ref.~\cite{porras_verstraete_06} for PBC, \emph{eMPS} to accentuate the difference.

In our experience, there are two main limitations of eMPS.  The first is that it is difficult to construct good variational guesses for the excited states in contrast to the ground state where the infinite size MPS algorithm~\cite{white_92,McCulloch_08} is applicable.  Here, the usefulness of the variance becomes readily apparent, as the discarded weight can be $10^{-12}$ or less while the variance is of order $10^{-2}$ in early sweeps.  The second is that that the area law considerations which make MPS algorithms so practical for ground states do not in general apply to bulk eigenstates, and so the bond dimension required to accurately represent a general eigenstate may be exponential in the system size, rendering eMPS inapplicable.

The ability to find excited states is useful in many contexts.  It provides access to the dynamical gap for determining the location of second order quantum phase transitions~\cite{sachdev1999} and Kibble-Zurek scalings, even when the gap is not between different symmetry sectors.  It can help in understanding the structure of conformal field theories by providing access to the primary scaling fields~\cite{Alcaraz_Berganza_11}.  Excited states yield the structure function and other dynamic response functions of low-lying excitations.  Such response functions are of great use for comparing to experimental measurements.  Finally, by considering more complex functionals such as $\langle \psi|(\hat{H}-\epsilon)^2|\psi\rangle-\lambda\langle \psi|\psi\rangle$ for minimization, one can determine level spacing statistics in a desired energy range for systems much larger than are amenable to exact diagonalization.  Such studies are immensely useful in discussions of integrability and quantum chaos, as well as investigations of the thermalization hypothesis~\cite{rigol2008,rigol2009c,gogolin2011}.

\section{Calculation of observables}
\label{sec:Obs}
We now turn to how we can extract information from a state expressed as an MPS.  We do so by the expectation values of Hermitian operators, or \emph{observables}.  We will demonstrate how to compute observables of three different types: local observables, two-point correlation functions, and general MPOs.

We define a local observable as an operator which acts only on the Hilbert space of a single site: $\hat{O}^{\left[k\right]}=\sum_{i,i'}O^{\left[k\right]}_{ii'}|i\rangle\langle i'|$.  If this site corresponds to the orthogonality center of $|\psi\rangle$ then the expectation value reduces to
\begin{eqnarray}
\langle \psi|\hat{O}^{\left[k\right]}|\psi\rangle&=\sum_{i_k i_k'}\mathrm{Tr}\left({A^{\left[k\right]i_k}}^{\dagger}O^{\left[k\right]}_{i_ki_k'}A^{\left[k\right]i_k'}\right)\, .
\end{eqnarray}
The overall scaling for fixed site index $k$ is $\mathcal{O}\left(\chi^2d^2\right)$.

A two-point correlation function is an expectation value of the form $\langle {\hat{O}^{\left[q\right]\dagger}}\hat{O}^{\left[r\right]}\rangle$ where we take $q<r$ without loss of generality.  If the orthogonality center of the MPS, $k$, lies within the range $q\le k\le r$, then we can evaluate the expectation value using only the tensors in this range.  The most efficient way to proceed is first to form the matrix
\begin{eqnarray}
R^{\left[j\right]}&=\sum_{i_r i_r'}A^{\left[r\right]i_r'}O^{\left[r\right]}_{i_ri_r'}{A^{\left[r\right]i_r}}^{\dagger}\, ,
\end{eqnarray}
recursively generate
\begin{eqnarray}
R^{\left[r-\ell\right]}&=\sum_{i}A^{\left[r-\ell\right]i}R^{\left[r-\ell+1\right]}{A^{\left[r-\ell\right]i}}^{\dagger}
\end{eqnarray}
for $\ell=1,\dots,r-q+1$, and then evaluate
\begin{eqnarray}
\langle {\hat{O}^{\left[i\right]\dagger}}\hat{O}^{\left[j\right]}\rangle &=\sum_{i_q,i_q'}\mathrm{Tr}\left(O^{\left[q\right]}_{i_qi_q'}A^{\left[i\right]i_q'}R^{\left[q+1\right]}{A^{\left[q\ell\right]i_q}}^{\dagger}\right)\,.
\end{eqnarray}
For fixed $q$ and $r$, the algorithm scales as $\mathcal{O}\left(\chi^3d+\chi^2d^2\right)$.

To compute the expectation of a general many-body observable $\hat{O}$ expressed as an MPO we start from the right (left) boundary and follow the recursion Eq.~(\ref{eq:LLeftrecur}) (Eq.~(\ref{eq:RRightrecur})) all the way to the opposite boundary, at which point the remaining $1\times 1\times 1$ tensor is the expectation value.  The overall scaling is $\mathcal{O}\left(\chi^3d\chi_O+\chi^2d^2M\left(\chi_O\right)\right)$ where $\chi_O$ is the bond dimension of the MPO and $M\left(\chi_O\right)$ is the number of nonzero elements in the MPO, which usually scales as $\chi_O$.

Entanglement measures such as the bond entropy
\begin{eqnarray}
S_{j}\equiv -\mathrm{Tr}\hat{\rho}_{j}\log \hat{\rho}_j\, ,\\
\hat{\rho}_{j}\equiv \mathrm{Tr}_{i<j+1}|\psi\rangle\langle \psi|\, ,
\end{eqnarray}
can be calculated from the singular values $\mathbf{\Sigma}$ of the MPS tensor $A^{\left[j\right]i}_{\alpha\beta}$ as
\begin{eqnarray}
S_j&=-\sum_{\gamma}\Sigma_{\gamma}^2\log \Sigma_{\gamma}^2
\end{eqnarray}
when this tensor is the orthogonality center.  These singular values are computed automatically as part of the algorithm to shift the orthogonality center, see Sec.~\ref{sec:MPS}.

\section{Time evolution with MPSs}
\label{sec:TE}

We now turn our attention to a variational solution of the time-dependent Schr\"odinger equation
\begin{eqnarray}
i\hbar \frac{\partial}{\partial t}|\psi\left(t\right)\rangle&=\hat{H}\left(t\right)|\psi\left(t\right)\rangle
\end{eqnarray}
using MPSs.  The general strategy is to find some representation of the propagator over some time interval $\left[t,t+\delta t\right]$, $\hat{U}(t,t+\delta t)$, and variationally optimize the functional
\begin{eqnarray}
\label{eq:Ufunctional}\left||\psi\left(t+\delta t\right)\rangle-\hat{U}\left(t,t+\delta t\right)|\psi\left(t\right)\rangle\right|^2
\end{eqnarray}
with respect to the MPS tensors of $|\psi(t+\delta t)\rangle$.  Several complications arise in this case which were not present in the earlier algorithms.  The first practical consideration is that the MPO form of the propagator may be difficult and very expensive to calculate.  The second difficulty is more physical; the time-dependent state following a global quench of a Hamiltonian parameter has entanglement which generally grows linearly in time~\cite{Calabrese_Cardy_05}.  This causes the bond dimension $\chi$ to grow exponentially in time, and so there is some finite time where an MPS simulation will exhaust the available computational resources.  However, many important questions regarding non-equilibrium dynamics can still be answered by considering moderately sized systems and short times.  In addition, consideration of a situation in which the Hamiltonian changes only locally can greatly increase the accessible system sizes and simulation times~\cite{carr2009l,carr2009m,carr2011j}.

The most common approach to time evolution for MPSs is to use the Suzuki-Trotter expansion
\begin{eqnarray}
\exp\left[-i\delta t\left(\sum_{n=1}^{L-1}\hat{H}_n\right)\right]&=\prod_{n=1}^{L-1}\exp\left(-\frac{i\delta t}{2}\hat{H}_n\right)\prod_{n=L-1}^{1}\exp\left(-\frac{i\delta t}{2}\hat{H}_n\right)+\mathcal{O}\left(\delta t^3\right)
\end{eqnarray}
or its higher order variants to construct a series of two-site propagators which can be constructed and applied easily.  This is the basis of the equivalent~\cite{Schollwoeck_11} TEBD~\cite{vidal_03} and tDMRG~\cite{white2004,daley2004} algorithms.  Here $\hat{H}_n$ is the nearest-neighbor bond Hamiltonian acting on sites $n$ and $n+1$.  This approach is no longer viable when the Hamiltonian has longer-ranged terms, and attempts to accommodate such longer-ranged terms often exhibit poor scaling~\cite{Stoudenmire_White_10,Schachenmayer_Lesanovsky_10} and require Hamiltonian-specialized implementation, resulting in inefficient, sometimes prohibitively inefficient code.  Krylov-based time evolution, which will form the basis for our approach, has been considered in both DMRG~\cite{Schmitteckert_04} and MPS~\cite{Garcia-Ripoll_06} variants for the time-independent case.  We note that the latter approach has been used~\cite{Garcia-Ripoll_06,Keilmann_GarciaRipoll_08} to study time dependent systems, but this necessitated very small time steps set by the rate of change of the Hamiltonian in order to provide accurate results.  Our approach generalizes the latter method to the time-dependent case where the error is independent of the rate of change of the Hamiltonian and demonstrates how the algorithm can be formulated entirely in terms of FSA rules for MPOs.  

\subsection{Commutator-free Magnus Expansions}
\label{sec:CFME}
The propagator of a general time-dependent Hamiltonian which does not commute with itself at different times is given as a time-ordered exponential
\begin{eqnarray}
\hat{U}\left(t,t+\delta t\right)\equiv \mathcal{T}\left[\exp\left(\int_t^{t+\delta t}dt'\hat{H}\left(t'\right)\right)\right]
\end{eqnarray}
whose most well-known form is the Dyson series
\begin{eqnarray}
\hat{U}\left(t,t+\delta t\right)=\hat{1}+\int_t^{t+\delta t}dt_1\hat{H}\left(t_1\right)+\int_t^{t+\delta t}dt_1\int_t^{t+\delta t}dt_2\hat{H}\left(t_1\right)\hat{H}\left(t_2\right)+\dots
\end{eqnarray}
This formulation of the propagator is not convenient numerically, as the Dyson series is an asymptotic series and so it can be difficult to determine an appropriate criteria for termination of the series.  Furthermore, keeping only a finite number of terms in the Dyson series does not preserve the Lie group structure of the propagator; that is, the finite approximation is not unitary.  An alternative approach which produces unitary approximations to the propagator was given by Magnus~\cite{Magnus_54} who used the ansatz
\begin{eqnarray}
\hat{U}\left(0,t \right)=\exp\left(-it\hat{\Omega}\left(t\right)\right)
\end{eqnarray}
to define the \emph{Magnus series}
\begin{eqnarray}
\label{eq:MagnusSeries}\hat{\Omega}\left(t\right)=\sum_{n=1}^{\infty}\hat{\Omega}_n\left(t\right)
\end{eqnarray}
where the $n^{\mathrm{th}}$ term is of order $t^{n}$ in the sense that its power series in $t$ starts with $t^n$.  The term $\hat{\Omega}_n\left(t\right)$ involves $n$ nested integrations over $n-1$ nested commutators of $\hat{H}\left(t\right)$ at different times.  Explicitly, the first few terms are:
\begin{eqnarray}
\hspace*{-2.5cm}\hat{\Omega}_1\left(t\right)=\int_0^tdt_1\hat{H}\left(t_1\right)\, ,\\
\hspace*{-2.5cm}\hat{\Omega}_2\left(t\right)=\frac{1}{2}\int_0^tdt_1\int_0^{t_1}dt_2\left[\hat{H}\left(t_1\right),\hat{H}\left(t_2\right)\right]\, ,\\
\hspace*{-2.5cm}\hat{\Omega}_3\left(t\right)=\frac{1}{6}\int_0^tdt_1\int_0^{t_1}dt_2\int_0^{t_2}dt_3\left(\left[\hat{H}\left(t_1\right),\left[\hat{H}\left(t_2\right),\hat{H}\left(t_3\right)\right]\right]+\left[\left[\hat{H}\left(t_1\right),\hat{H}\left(t_2\right)\right],\hat{H}\left(t_3\right)\right]\right)\, .
\end{eqnarray}
While approximations obtained from truncating the series yield unitary propagators, these expressions are still formidable numerically, involving nested commutators and multidimensional integrals.  The commutators pose a special difficulty for MPOs, as exact multiplication of MPOs involves multiplication of the bond dimensions of the MPOs and hence the algorithm scales exponentially in the number of terms kept in the series.  Optimization algorithms which attempt to variationally shrink the bond dimension of an MPO sum or product such as those proposed in Ref.~\cite{Froewis_Nebendahl_10} may also be used, but these become numerically unstable for large systems, and when MPOs are subtracted as in commutators large cancellations can cause these algorithms to become stuck far from the variational optimum.

Hence, rather than work directly with the Magnus series, Eq.~(\ref{eq:MagnusSeries}), we start from ans\"atze of the form
\begin{eqnarray}
\hat{U}\left(t,t+\delta t \right)=\prod_{i=1}^s \exp\left(-i\delta t\hat{\Omega}_i\right)
\end{eqnarray}
where each one of the $\hat{\Omega}_i$ is a linear combination of $\hat{H}$ at different times in the interval $\left[t,t+\delta t\right]$, and require that our ansatz matches the Magnus expansion (equivalently, the full propagator) up to order $\delta t^{N+1}$.  We will call such an ansatz a commutator-free Magnus expansion (CFME)~\cite{Blanes_Moan_06,Blanes_Casas_09}.  This ansatz has a number of features which make it desirable for our purposes.  It is exactly unitary and so the norm is conserved.  Also, provided that we consider the case where only the Hamiltonian parameters change in time and the operators are time-independent, the sums of the Hamiltonian at different times can be represented exactly as an MPO using the rules of Sec.~\ref{sec:MPO}.  Thus, the need for complex operations with MPOs vanishes.  Finally, because the ansatz takes into account the time dependence of the Hamiltonian explicitly, the time step is not necessarily fixed by the rate of variation of the Hamiltonian, allowing for more coarse stepping in time with fixed error.

Following Ref.~\cite{Alvermann_Fehske_11}, the procedure for generating an $N^{\mathrm{th}}$-order CFME is to expand the function $H\left(t\right)$ in terms of (shifted) Legendre polynomials $P_n$, 
\begin{eqnarray}
\label{eq:Hn}\hat{H}_n&=\left(2n-1\right)\delta t\int_0^1dx\hat{H}\left(t+x\delta t\right)P_{n-1}\left(x\right)\, .
\end{eqnarray}
The orthogonality properties of the Legendre polynomials allow the nested integration to be done exactly, leaving a series of nested commutators of the $\hat{H}_n$.  Working in a Hall basis~\cite{Munthe_Owren_99}, this series of commutators is matched with the original Magnus expansion to yield the order conditions $f_{i,n}$ such that
\begin{eqnarray}
\hat{\Omega}_i=\sum_{n=1}^{N}f_{i,n}\hat{H}_n\, .
\end{eqnarray}
We note that these order conditions are independent of $\hat{H}\left(t\right)$ by construction, and so are set by the choice of CFME alone.  As we only require the result to be valid to order $\delta t^{N+1}$, the integration required for the coefficients $\hat{H}_n$ may be performed using Gauss-Legendre quadrature of order $N/2+1$.  The end result of the analysis is that an $N^{\mathrm{th}}$ order expansion with $s$ exponentials may be written as
\begin{eqnarray}
\label{eq:CFME}\hat{U}\left(t+\delta t,t\right)=e^{-i\delta\hat{\Omega}_1}\dots e^{-i\delta\hat{\Omega}_s}\\
\label{eq:Omegags}\hat{\Omega}_i= \sum_{m=1}^{N/2+1}g_{i,m}\hat{H}\left(t+x_m\delta t\right)\\
g_{i,m}=w_m\sum_{n=1}^{N/2+1}\left(2n-1\right)P_{n-1}\left(x_m\right)f_{i,n}
\end{eqnarray}
where $x_m$ and $w_m$ are the points and weights for Gauss-Legendre quadrature~\cite{press1993}.  In this work we use a fourth order expansion with three exponentials ($N=4$, $s=3$).  The optimal order conditions for this case, obtained in Ref.~\cite{Alvermann_Fehske_11}, are
\begin{eqnarray}
f_{1,1}=11/40\, ,\;\; f_{1,2}=20/87\, ,\;\; f_{1,3}=7/50\, ,\;\; f_{2,1}=9/20 \, ,\;\; f_{2,3}=-7/25\, ,
\end{eqnarray}
with $f_{s-i+1,n}=\left(-1\right)^{n+1}f_{i,n}$.  Order conditions for higher order expansions may also be found in Ref.~\cite{Alvermann_Fehske_11}.

We now consider that our time-dependent Hamiltonian MPO is constructed from a set of FSA rules $\left\{\mathcal{R}_p\left(\left\{\hat{O}^{p_1},\dots,\hat{O}^{p_n}\right\},\left\{h^p\left(t\right)\right\},w^p\right)\right\}$ in which the OA and the weights are chosen without loss of generality to be time-independent.  In this case, the expansion Eq.~(\ref{eq:Hn}) is applied individually to each Hamiltonian parameter $h^p\left(t\right)$, resulting in the parameters $\left\{h^p_n\right\}$.  Now, because of the canonical decomposition of Sec.~\ref{sec:MPO}, the MPO forms of $\hat{\Omega}_i$ from Eq.~(\ref{eq:Omegags}) at time $t$ can be constructed exactly using the FSA rule set $\left\{\mathcal{R}_p\left(\left\{\hat{O}^{p_1},\dots,\hat{O}^{p_n}\right\},\left\{\sum_{m=1}^{N/2+1}g_{i,m}h^p\left(t+x_m\delta t\right)\right\},w^p\right)\right\}$.  We note that each one of these operators has the same bond dimensions as the original Hamiltonian, and the updates of operators $\hat{\Omega}_i$ at each time step can be done in $\mathcal{O}\left(L\chi_O\right)$ time, which is essentially negligible.

The fact that we can construct the $\hat{\Omega}_i$ from the same FSA rules as the Hamiltonian implies that our CFME ansatz is equivalent to evolving our system according to piecewise constant Hamiltonians of the same form but with differing Hamiltonian parameters.  Additionally, as also occurs in high-order Suzuki-Trotter expansions, evolution backwards in time may occur.  Finally, we note that even terms in the Hamiltonian whose parameters do not vary in time have their magnitude altered by Eq.~(\ref{eq:Omegags}), as $\sum_{m}g_{i,m}\ne 1$ in general.

\subsection{Krylov subspace propagation}
\label{sec:Krylov}
Using the CFME Eq.~(\ref{eq:CFME}) we never need to explicitly form an MPO representation of the propagator provided we can find an MPS representation of the exponential of an MPO applied to an MPS.  We find such a representation from minimizing functionals of the form
\begin{eqnarray}
\left||\phi\rangle-\exp\left(-i\delta t\hat{\Omega}\right)|\psi\rangle\right|^2\, ,
\end{eqnarray}
where, importantly, $\hat{\Omega}$ has a known MPO representation.  We do so by forming a Krylov subspace approximation to the exponential~\cite{Saad_92} in which the Krylov vectors are represented as MPSs.  Specifically, we do so via the Lanczos algorithm for the matrix exponential, which can be stated as
\begin{enumerate}
\item \emph{Input}:  Input $\hat{\Omega}$ in MPO form, $|\psi\rangle$ in MPS form, and a tolerance $\epsilon$ for truncating the recursion.
\item \emph{Initialize}: Set $\beta_0=\langle \psi|\psi\rangle=1$ and $|r\rangle=|\psi\rangle$.
\item \emph{Iterate}: For $j=1,2,\dots$ until convergence
\begin{enumerate}
\item $|v_j\rangle=|r\rangle/\beta_{j-1}$.
\item $|r\rangle=\hat{\Omega}|v_j\rangle$.
\item $\alpha_j=\langle v_j|r\rangle$.
\item Orthogonalize $|r\rangle$ against $|v_j\rangle$ and $|v_{j-1}\rangle$.
\item Re-orthogonalize $|r\rangle$ against all $v_k$, $k\le j$ if necessary.
\item $\beta_j=\langle r|r\rangle$
\item Form the matrix exponential of $T^{\left(j\right)}$, $U^{\left(j\right)}$, and obtain $c^{\left(j\right)}=U^{\left(j\right)}_{1:j,1}.$
\item Test for convergence.
\end{enumerate}
\item \emph{Finalize:} Set $|\phi\rangle=\sum_{i=1}^{j}c^{\left(j\right)}_i|v_i\rangle$
\end{enumerate}
Here $T^{\left(j\right)}$ is the symmetric tridiagonal matrix with the $\alpha_i$, $1\le i\le j$ on the diagonal and $\beta_i$, $1\le i\le j-1$ on the superdiagonal.  It is important to use a matrix exponentiation method which produces a unitary matrix to machine precision in order to not lose the Lie group structure.  Because of the small linear dimensions of the matrix $T^{\left(j\right)}$, exponentiation by direct diagonalization is practical.  An \emph{a posteriori} estimate for convergence of the Lanczos recursion is that $\left|2\beta_{j-1}c_{j}^{\left(j\right)}\right|<\epsilon$, where $\epsilon$ is the tolerance~\cite{Saad_92}.  This can be compared with residual estimates in the ordinary Lanczos algorithm for finding eigenvalues.  A rigorous bound on the approximation $\left||\psi\rangle_{\mathrm{krylov}}-|\psi\rangle\right|\le 12\exp\left\{-\frac{\left(\rho \delta t\right)^2}{16 n}\right\}\left(\frac{ e\rho \delta t}{4n}\right)^n$ can be established~\cite{Hochbruck_Lubich_97} when $n\ge \rho \delta t/2$ with $n$ the number of Lanczos vectors and $\rho=\left|E_{\mathrm{max}}-E_{\mathrm{min}}\right|$ the spectral width of $\hat{\Omega}$.  This estimate shows that for typical tolerances $\epsilon=10^{-6}$ to $10^{-10}$, $6$ to $20$ Lanczos vectors suffice.

As stated before, MPSs do not form a vector space and so the multiplication by $\hat{\Omega}$, the orthogonalization, and the final summation cannot be done exactly while keeping the bond dimension of our MPS fixed.  However, just as with the eigenstate search, we can devise variational algorithms for these three operations which are iteratively performed until a desired tolerance is reached and use this tolerance to bound the bond dimension of our time-evolved MPS.  We begin by briefly reviewing the standard algorithm~\cite{Schollwoeck_11} for finding the optimal MPS $|\phi\rangle$ representing a sum of MPSs $\sum_{k=1}^{N}c_k|\psi_k\rangle$ to a given tolerance, as the other algorithms are similar but more complex.  In this case we have a set of $\mathcal{LR}$ overlaps defined between our variational state $\langle\phi|$ and the states $|\psi_k\rangle$ as in Eqs.~(\ref{eq:Leftrecur}) and (\ref{eq:Rightrecur}).  We now sweep through the lattice and make the replacement
\begin{eqnarray}
\label{eq:fit}A^{\left[j\right]}\dots A^{\left[j+s-1\right]}&=\sum_kc_kF^{\left[j\right]\left\{k\right\}}\, ,
\end{eqnarray}
where the $F^{\left[j\right]\left\{k\right\}}$ are formed as in Eq.~(\ref{eq:Fdef}), see also Fig.~\ref{fig:ESdiagrams}(b).  The orthogonality center of $|\phi\rangle$ is then shifted, the $\mathcal{LR}$ overlaps updated, and sweeping continued.  Convergence can be monitored via
\begin{eqnarray}
\left||\phi\rangle-\sum_kc_k|\psi_k\rangle\right|^2&=&\mathrm{Tr}\left[\left(A^{\left[j\right]}\dots A^{\left[j+s-1\right]}\right)^{\dagger}A^{\left[j\right]}\dots A^{\left[j+s-1\right]}\right]+1\\
\nonumber&&-2\sum_k\mathrm{Re}\,\mathrm{Tr}\left[c_k\left(A^{\left[j\right]}\dots A^{\left[j+s-1\right]}\right)^{\dagger} F^{\left[j\right]\left\{k\right\}}\right]<\epsilon\,,
\end{eqnarray}
with $\mathrm{Re}$ denoting the real part.  Because we do not have to solve an eigenequation at each iteration, this algorithm is often much less costly than the iterative eigenstate search.  Also, when we have that the coefficient vector $c$ and all of the $\left\{|\psi_k\rangle\right\}$ have length 1, we can normalize the state $|\phi\rangle$ at the end of the calculation if required.

The algorithm to variationally fit an MPS to $\hat{\Omega}|\psi\rangle$ is similar.  In this case we have a set of $LR$ overlaps defined via the recursions
\begin{eqnarray}
L^{\left[1\right]}_{\kappa \alpha\alpha'}&=\delta_{\alpha,1}\delta_{\alpha',1}\delta_{\kappa,1}\, ,\\
L^{\left[\ell+1 \right]}_{\kappa\alpha\alpha'}&=\left[\sum_{\gamma' i'}\left[\sum_{i \kappa'}\left[\sum_{\gamma}{A^{\left[\ell\right]i}_{\gamma\alpha}}^{\star}L^{\left[\ell\right]}_{\kappa' \gamma\gamma'}\right]W^{\left[ \ell\right] ii'}_{\kappa'\kappa}\right]B^{\left[\ell\right]i'}_{\gamma'\alpha'}\right]\, ,\\
R^{\left[L+1\right]}_{\kappa \beta'\beta}&=\delta_{\beta,1}\delta_{\beta',1}\delta_{\kappa,1}\, ,\\
R^{\left[\ell \right]}_{\kappa\beta'\beta}&=\left[\sum_{\gamma'i'}B^{\left[\ell\right]i'}_{\beta'\gamma'}\left[\sum_{\kappa'i}W_{\kappa\kappa'}^{\left[\ell\right]ii'}\left[\sum_{\gamma}R_{\kappa'\gamma'\gamma}^{\left[\ell+1\right]}{A^{\left[\ell\right]i}_{ \beta\gamma}}^{\star}\right]\right]\right]\, ,
\end{eqnarray}
where the MPS tensors of $|\phi\rangle$ are denoted by $A$ and those of $|\psi\rangle$ denoted by $B$.  We now sweep through the lattice and make the replacement
\begin{eqnarray}
A^{\left[j\right]}\dots A^{\left[j+s-1\right]}&=\hat{H}_{\mathrm{eff}}^{\left[j\right]}B^{\left[j\right]}\dots B^{\left[j+s-1\right]}
\end{eqnarray}
where the effective Hamiltonian is formed from the $LR$ overlaps as in Eq.~(\ref{eq:EH}).  Convergence can be monitored via 
\begin{eqnarray}
\label{eq:Hcriterion}\left||\phi\rangle-\hat{H}|\psi\rangle\right|^2&=&\mathrm{Tr}\left[\left(A^{\left[j\right]}\dots A^{\left[j+s-1\right]}\right)^{\dagger}A^{\left[j\right]}\dots A^{\left[j+s-1\right]}\right]+\langle \psi|\hat{H}^2|\psi\rangle\\
\nonumber&&-2\mathcal{R}\mathrm{Tr}\left[\left(A^{\left[j\right]}\dots A^{\left[j+s-1\right]}\right)^{\dagger}\hat{H}_{\mathrm{eff}}B^{\left[j\right]}\dots B^{\left[j+s-1\right]}\right]<\epsilon\, ,
\end{eqnarray}
Here, $\langle \psi|\hat{H}^2|\psi\rangle$ can be computed in a manner similar to the variance, and need only be computed once at the beginning of the calculation.  We have also assumed that the block of tensors in Eq.~(\ref{eq:Hcriterion}) contains the orthogonality center.  Again, this algorithm is often much less costly than the iterative eigenstate search.

We now turn to steps (iii)(d) and (iii)(e) of our algorithm.  Step (iii)(d) is usually stated for ordinary vector spaces as
\begin{eqnarray}
|r\rangle&=|r\rangle-\alpha_j|v_j\rangle-\beta_{j-1}|v_{j-1}\rangle\, ,
\end{eqnarray}
as $\alpha_j=\langle v_j|r\rangle$ and $\beta_{j-1}=\langle v_{j-1}|r\rangle$ and so this is equivalent to classical Gram-Schmidt orthogonalization.  Hence, we could implement step (iii)(d) by using the fitting algorithm Eq.~(\ref{eq:fit}) to find the MPS closest to $|r\rangle-\alpha_j|v_j\rangle-\beta_{j-1}|v_{j-1}\rangle$.  However, we have found that the following algorithm, which is closely related to eMPS, often converges more quickly and also is applicable to step (iii)(e).  In our method we look for the optimal MPS $|\phi\rangle$ representing $|\psi\rangle$ but also subject to the constraints that $\langle \phi|\psi_k\rangle=0$ for some set $\left\{|\psi_k\rangle\right\}$.  We start by copying the state $|\psi\rangle$ to a variational ansatz $|\phi\rangle$.  We then construct overlaps between the state $\langle \phi|$ and $|\psi\rangle$, which we call $LR$\footnote{In this initialization all of the $LR$ are Kronecker deltas provided that $|\psi\rangle$ has an orthogonality center.} and a set of overlaps between $\langle \phi|$ and $|\psi_k\rangle$, which we call $\mathcal{LR}$.  We then sweep through the lattice and make the replacement
\begin{eqnarray}
A^{\left[j\right]}\dots A^{\left[j+s-1\right]}&=\sum_{\gamma_1\dots \gamma_{s+1}} {L}^{\left[j\right] i_j}_{\alpha\gamma_1}B^{\left[j\right]i_j}_{\gamma_1\gamma_2}\dots B^{\left[j+s-1\right]i_{j}}_{\gamma_s\gamma_{s+1}} {R}^{\left[j+s\right]}_{\gamma_{s+1}\beta}\, ,
\end{eqnarray}
with $B$ the MPS tensors of $|\psi\rangle$ and $A$ the MPS tensors of $|\phi\rangle$.  We then apply the projector into the space orthogonal to the $|\psi_k\rangle$ by constructing the set $\left\{G^{\left[j\right]\left\{\mu\right\}}\right\}$, $\mu=1,\dots,p$, as in Eq.~(\ref{eq:Gconst}) and performing
\begin{eqnarray}
A^{\left[j\right]}\dots A^{\left[j+s-1\right]}&=A^{\left[j\right]}\dots A^{\left[j+s-1\right]}-\mathrm{Tr}\left({G^{\left[j\right]\left\{\mu\right\}}}^{\dagger}A^{\left[j\right]}\dots A^{\left[j+s-1\right]}\right){G^{\left[j\right]\left\{\mu\right\}}}\, ,
\end{eqnarray}
for $\mu=1,\dots, p$.  Using the fitting algorithm of Eq.~(\ref{eq:fit}) corresponds to replacing the $\mathcal{LR}$ overlaps, which are between the \emph{variational} state $\langle \phi|$ and the set $\left\{|\phi_k\rangle\right\}$, with a set of $\mathcal{LR}$ overlaps between the state $\langle \psi|$ and the set $\left\{|\phi_k\rangle\right\}$.  Our algorithm, which amounts to fitting followed by modified Gram-Schmidt, uses information about the distance between the variational state and those to be orthogonalized against to determine operations, and hence often converges more quickly and is more stable.  Convergence can be monitored by ensuring that $\langle \phi|\psi_k\rangle$ are orthogonal to a precision $\epsilon$ via
\begin{eqnarray}
\left|\mathrm{Tr}\left({A^{\left[\ell\right]}}^{\dagger}F^{\left[\ell\right]\left\{k\right\}}\right)\right|\le \epsilon\, .
\end{eqnarray}
If one requires additional truncation of the bond dimension, one can switch to the ordinary fitting algorithm Eq.~(\ref{eq:fit}) at this point, using a new variational state $|\zeta\rangle$ to fit to $|\phi\rangle$.

We now pause to consider the sources of error in the time-propagation routine.  First, because the CFME expansion Eq.~(\ref{eq:CFME}) is only of order $\delta t^{N+1}$, the error incurred in using this form of the propagator is $\epsilon_{\mathrm{CFME}}=ct_{\mathrm{final}}\delta t^N$, where $t_{\mathrm{final}}$ is the final time reached.  Thus, as the final time desired becomes longer, smaller time steps should be taken in order to keep the error fixed.  The coefficient $c$ can be determined by using this known scaling and decreasing the time step by a constant factor.  Factors in the range $2^{1/N}$ to $3^{1/N}$ are practical.  This strategy can also be used to devise adaptive time-step applications such as those used widely in ordinary differential equation solvers.  Next, there is the error $\epsilon_{\mathrm{Krylov}}$ incurred in the Krylov subspace approximation to the exponential.  As discussed above, this error can be minimized by adding more and more Lanczos vectors.  This error also involves the time step $\delta t$, and so when determining the CFME expansion error constant $c$ one should be careful that $\epsilon_{\mathrm{Krylov}}<\epsilon_{\mathrm{CFME}}$.  Finally, there are errors resulting from the variational fitting of MPSs in steps (iii)(b), (iii)(d), and (iii)(e) of the Lanczos algorithm for the matrix exponential.  These can be reduced by lowering the variational tolerances, but this is done typically at the expense of a larger bond dimension $\chi$ and hence a slower algorithm and more memory usage.

\subsection{Simulation Protocol}
\label{sec:protocol}
We are now in a position to devise a complete, generic protocol for the time evolution of a 1D quantum system.
\begin{enumerate}
\item \emph{Input}: Input an operator alphabet and a set of FSA rules defining the Hamiltonian MPO.  Input the functional forms $\left\{h_p\left(t\right)\right\}$ of its Hamiltonian parameters, a final time desired $t_{\mathrm{final}}$ and a time-step $\delta t$.  Input tolerances $\left\{\epsilon\right\}$ for variational ground state search and time evolution.
\item \emph{Initial state preparation}: Find the ground state of the Hamiltonian using variational ground state search from Sec.~\ref{sec:vMPS}.  Alternatively, if a different initial state is desired, read in its MPS representation.  Set $|\psi\left(t=0\right)\rangle$ to be this state.
\item \emph{Measure}: Measure local observables and two-point correlation functions constructed from the OA as well as MPO observables constructed from their own FSA rule sets.
\item \emph{eMPS}: Use eMPS (Sec.~\ref{sec:eMPS}) to find excited states of the Hamiltonian, if desired.  Measure properties of the excited states.
\item \emph{tMPS}:  Set $t=0$.
\begin{enumerate}
\item For $i=s,\dots 1$, construct $\hat{\Omega}_i$ from Eq.~(\ref{eq:Omegags}) using the FSA rules and use the Krylov algorithm to apply the matrix exponentials of these operators in succession to $|\psi\left(t\right)\rangle$. Set $t=t+\delta t$.  
\item Measure $|\psi\left(t\right)\rangle$ if desired.
\item Continue (a) and (b) until $t_{\mathrm{final}}$ is reached.
\end{enumerate}
\end{enumerate}

Starting from a base set of rules such as the site, bond, exponential, decaying function, and finite function rules of Sec.~\ref{sec:MPO}, a single implementation can accommodate a vast range of systems based on the particular OA, rules, and quench protocols $\left\{h_p\left(t\right)\right\}$ used.  

\section{Case studies}
\label{sec:CaseStudies}
\subsection{Case Study: Ising model in a transverse field}
For our first case study we choose the paradigmatic Ising model in a transverse field, with Hamiltonian
\begin{eqnarray}
\hat{H}=-J\sum_{\langle i,j\rangle}\hat{\sigma}^z_i\hat{\sigma}^z_{i+1}-h\sum_{i}\hat{\sigma}^x_i\, .
\end{eqnarray}
Here $J$ is the coupling energy, $h$ is a transverse magnetic field, and the $\left\{ \hat{\sigma}_i\right\}$ are the Pauli spin operators on site i.  We choose this model because its dynamics are amenable to numerically exact study using the time-dependent Bogoliubov-de Gennes formalism (see \ref{sec:IsingExact} for a review) and so we were able to carefully check convergence of our results.  The statics have all been converged to eight digits, and the dynamics at all times to at least four digits.  Here we refer to convergence of local quantities such as the energy or density of defects.  Nonlocal quantities such as the bond entropy will not have this same accuracy, but numerical tests show that the qualitative behavior is unaffected.  We begin with a discussion of the statics.

\begin{figure}[t]
\begin{center}
\epsfxsize=0.8\textwidth  \leavevmode \epsfbox{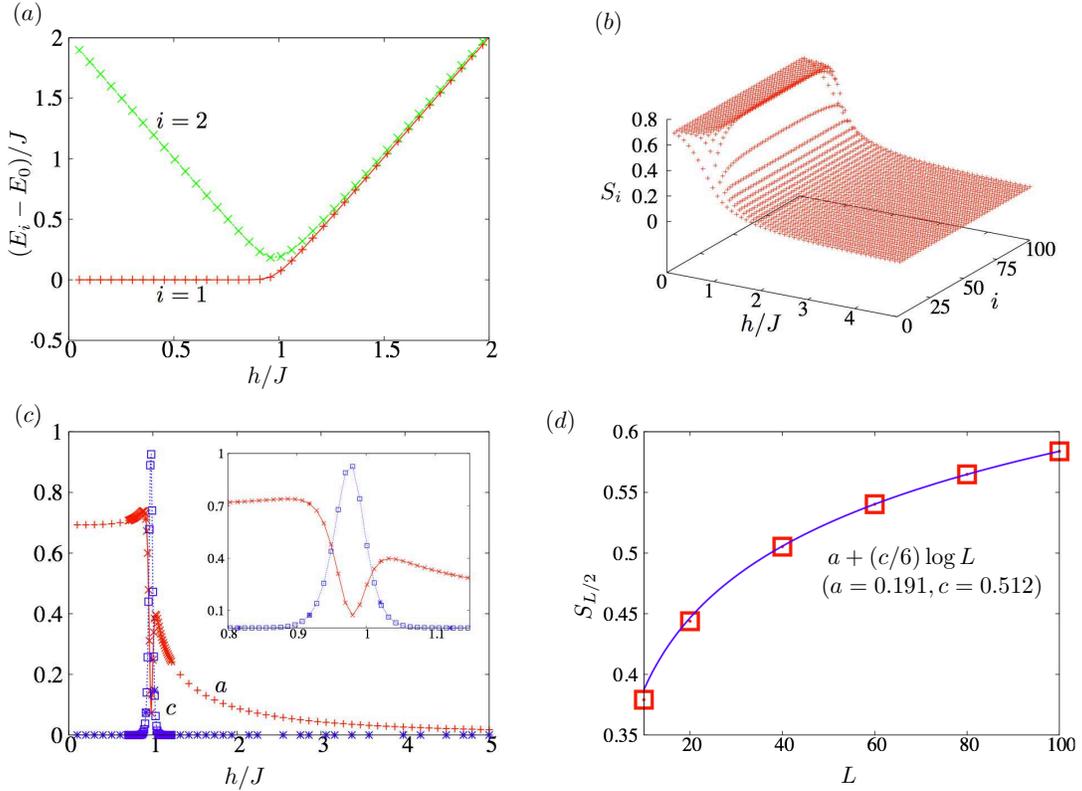}
\caption{  \label{fig:IsingStatics}  (color online) \emph{Statics of the Ising model.} (a) The gaps to the two-lowest-lying eigenstates, computed using eMPS.  (b) The bond entropy $S_i$ vs.~the site index $i$ and magnetic field $h$ for $L=100$ sites.  Note the increased curvature near the critical point $h=1$.  (c) The central charge $c$ (blue line) and bulk entanglement $a$ (red line) extracted from a fit to Eq.~(\ref{eq:finiteCC}) vs.~$h$ for $L=100$ sites, neglecting 30 sites at both boundaries.  The inset is a close-up of the critical region.  (d) The bond entropy of the central site $S_{L/2}$ vs.~$L$ at $h=1.0$ together with a fit extracting the central charge.}
\end{center}
\end{figure}

In Fig.~\ref{fig:IsingStatics}(a) we demonstrate the gaps from the ground state to the two lowest eigenstates, computed via eMPS.  The variances are smaller than the point size in this case.  The upper (lower) curve corresponds to even (odd) parity, while the ground state has even parity, where parity is defined as simultaneous inversion of all spins $P=\langle \prod_i\hat{\sigma}^x_i\rangle$.  Hence, the relevant gap for discussing the quantum phase transition is in fact the gap to the second excited state, shown in green.\footnote{The even parity gap for $h>1$ is in fact twice the demonstrated gap in green, but the essential piece is the closing of the gap at criticality.}  Both gaps close at criticality, and this can cause the first excited state returned by eMPS to be a mixture of these two states.  This will not affect the energies so long as the variance remains small, but it can affect other observable properties of the states.  There are two ways to remove this nuisance.  The first, most complex, and most preferable is to use an MPS representation in which the state is explicitly $\mathbb{Z}_2$ invariant~\cite{McCulloch_07}.  The second is to add a field coupling to the parity $-h_p\prod_i\hat{\sigma}^x_i$ to cause the relevant even-parity subspace to become lower in energy than the odd-parity subspace.  This operator is an MPO with bond dimension 1, the MPO equivalent of a product state.  The closing of the gap at the known critical point $h=1$ is linear in $1/L$, indicating a conformally invariant critical theory with dynamical critical exponent $z=1$.

We venture to determine the central charge of the critical theory by fitting to the Calabrese-Cardy formula in two ways.  In the first, we fit to the finite-size formula
\begin{eqnarray}
\label{eq:finiteCC}S_i&=\frac{c}{6}\log\left[\frac{2 L}{\pi} \sin\left(\frac{\pi i}{L}\right)\right]+a
\end{eqnarray}
for fixed $L$ and variable $i$, and in the second we fix $i$ at $L/2$ and fit $S_{L/2}$ to this formula for various $L$.  Near criticality, the presence of nonzero $c$ indicates a curvature of $S_i$, while in the gapped phases $S_i$ obeys an area law and is hence a constant apart from finite-size effects.  The bond entropy in the bulk approaches the correct limiting values of $\log 2$ as $h\to0 $ and $0$ as $h\to \infty$.  The first fit, shown in Fig.~\ref{fig:IsingStatics}(a), provides us with a clear indicator of the critical region by the spike in the central charge $c$.  However, the precise determination of $c$ for a finite size system in this case is noisy, likely due to strong finite-size effects.  Once we have narrowed down where the critical region is, the second fit, shown in  Fig.~\ref{fig:IsingStatics}(b), allows us to extract the anticipated value $c=1/2$ much more precisely.  If the same scaling analysis is attempted at a point which is not the critical point, the bond entropy saturates and $c\to 0$ as $L\to \infty$.  We can understand this as a large but finite correlation length $\xi$.  For $L/2<\xi$, the system appears to be conformal and we see scaling of the bond entropy with $L$.  For $L>\xi$ the bond entropy saturates and this scaling breaks down, indicating that the given region is not critical.  We note that in this analysis we have used no properties which are specific to this system e.g.~correlation functions of an order parameter to extract the critical behavior.

\begin{figure}[t]
\begin{center}
\epsfxsize=0.8\textwidth \leavevmode \epsfbox{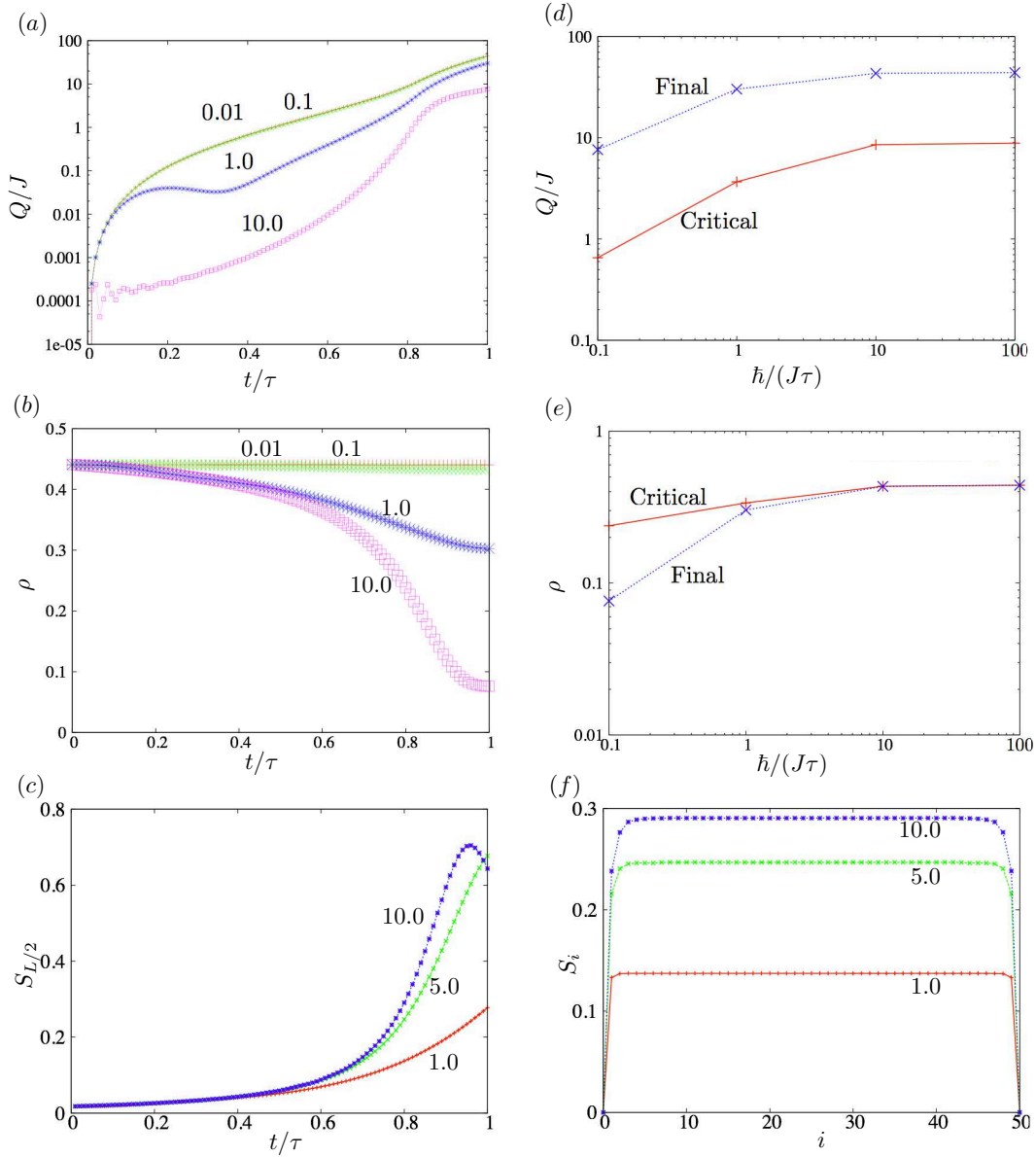}
\caption{  \label{fig:IsingDynamics}  (color online) \emph{Dynamics of the Ising model.} (a) The heat as a function of time shows a marked change in behavior as we transition past the critical time $t/\tau=4/5$.  The numbers indicate the value of $\tau J/\hbar$.  $\tau J/\hbar=0.1$ and $0.01$ are indistinguishable on the scale of this plot.  Both may be considered to be instantaneous.  (b) The density of defects as a function of time scales to zero as $\tau J/\hbar\to \infty$, in accordance with the KZH.  (c) The bond entropy of the central bond as a function of time approaches the limiting value $\log 2$ as the quench becomes more adiabatic.  For nearly adiabatic quenches, the bond entropy oscillates after the critical point.  (d) Scaling of the heat in the final and critical stages with the inverse quench time shows marked non-adiabatic processes occurring after passing the critical point.  (e) Scaling of the density of defects in the final and critical stages with the inverse quench time shows non-adiabatic processes after passing the critical point only for slow quenches.  (f) Snapshots of the bond entropy at the critical time demonstrate that the system is not generally close to its conformal ground state.}
\end{center}
\end{figure}

We now turn to the dynamics.  The Ising model has also been a subject of interest for dynamics as a testbed for the Kibble-Zurek hypothesis (KZH) that equilibrium properties determine nonequilibrium properties following a quench across a quantum critical point.  This was studied in Refs.~\cite{Zurek_Dorner_05, Dziarmaga_05, Cincio_Dziarmaga_07} using the quench protocol
\begin{eqnarray}
\label{eq:Quenchprotocol}h\left(t\right)/J&=5\left(1-\frac{t}{\tau}\right)\, ,\;\; 0\le t\le \tau\, .
\end{eqnarray}
A useful quantity for determining how non-adiabatic a particular quench is in this case is the density of defects
\begin{eqnarray}
\rho=\frac{1}{2L}\sum_i\left(1-\hat{\sigma}^z_i\hat{\sigma}^z_{i+1}\right)
\end{eqnarray}
which is the density of Bogoliubov quasiparticles at zero magnetic field.  In addition to the density of defects, universal scaling has also been predicted in the heat, or non-adiabatic part of the energy,
\begin{eqnarray}
Q\left(t\right)=\langle \psi\left(t\right)|\hat{H}\left(t\right)|\psi\left(t\right)\rangle-\langle \psi_{\mathrm{g.s.}}\left(t\right)|\hat{H}\left(t\right)|\psi_{\mathrm{g.s.}}\left(t\right)\rangle
\end{eqnarray}
where $|\psi_{\mathrm{g.s.}}(t)\rangle$ represents the instantaneous ground state of $\hat{H}(t)$.  In addition to these quantities, which are amenable to Bogoliubov-de Gennes analysis, we also compute the time-dependent bond entropy.  Our results are shown in Fig.~\ref{fig:IsingDynamics}.

We first discuss the heat, as shown in panels (a) and (d).  Panel (a) displays the heat as a function of time, and demonstrates a sharp change in the behavior of the system as we pass through the critical point.  This is especially true of the longest quenches.  In panel (d) we investigate the heat as a function of the quench rate both at the time when $h$ takes on its critical value, $t_c=4\tau/5$, and at the final time when $h=0$.  The large difference indicates that non-adiabatic processes continue after we have passed from the critical region back into the gapped ferromagnetic region.  Thus, the universal scaling of the heat may be difficult to determine if the critical point itself is not known sufficiently well.   We now move on to the density of defects as shown in panels (b) and (e).  In panel (b) we see that the density of defects at the final time decreases slowly to zero as $\tau\to\infty$; that is, when the quench becomes perfectly adiabatic.  This is in accordance with the KZH prediction.  Panel (e) demonstrates the large disparity between the density of defects at the critical time and the final time for all but the most rapid of quenches.  Finally, in panels (c) and (f) we investigate the bond entropy.  In panel (c) we see the bond entropy of the central bond as a function of time.  As the quench becomes more adiabatic, the entropy increases more towards the ferromagnetic limiting value of $\log 2$.  However, for very slow quenches, the bond entropy reaches this value before the end of the quench and then begins to oscillate.  In panel (f) we show the bond entropy as a function of the bond index at the critical time.  Bulk curvature such as that seen at criticality in Fig.~\ref{fig:IsingStatics}(d) is absent, indicating that the time-evolved state is still quite far from the conformal ground state.

\subsection{Case Study: Dipolar Ising chain}
In this section we go beyond exactly solvable models and study a dipolar Ising chain
\begin{eqnarray}
\hat{H}_{\mathrm{dip}}=-J\sum_{i<j;\left|j-i\right|\le 6}\frac{\hat{\sigma}^z_i\hat{\sigma}^z_j}{\left(j-i\right)^3}-h\sum_i\hat{\sigma}^x_i\, .
\end{eqnarray}
Such models are relevant to the study of ultracold molecules in optical lattices, where the dipole-dipole interaction falls away as $1/r^3$ with $r$ the distance between dipoles~\cite{Carr_Demille_09,Wall_Carr_10,Gorshkov_Manmana_11,Gorshkov_Manmana_11b,wallforthcoming}.  Here the cutoff $\left|j-i\right|\le 6$ represents a consistent order of approximation in going from a Hubbard-type model with dipolar interactions to a spin model.  We stress that all results obtained in this section were obtained using the same implementation as the last section.

\begin{figure}[t]
\begin{center}
\epsfxsize=0.8\textwidth \leavevmode \epsfbox{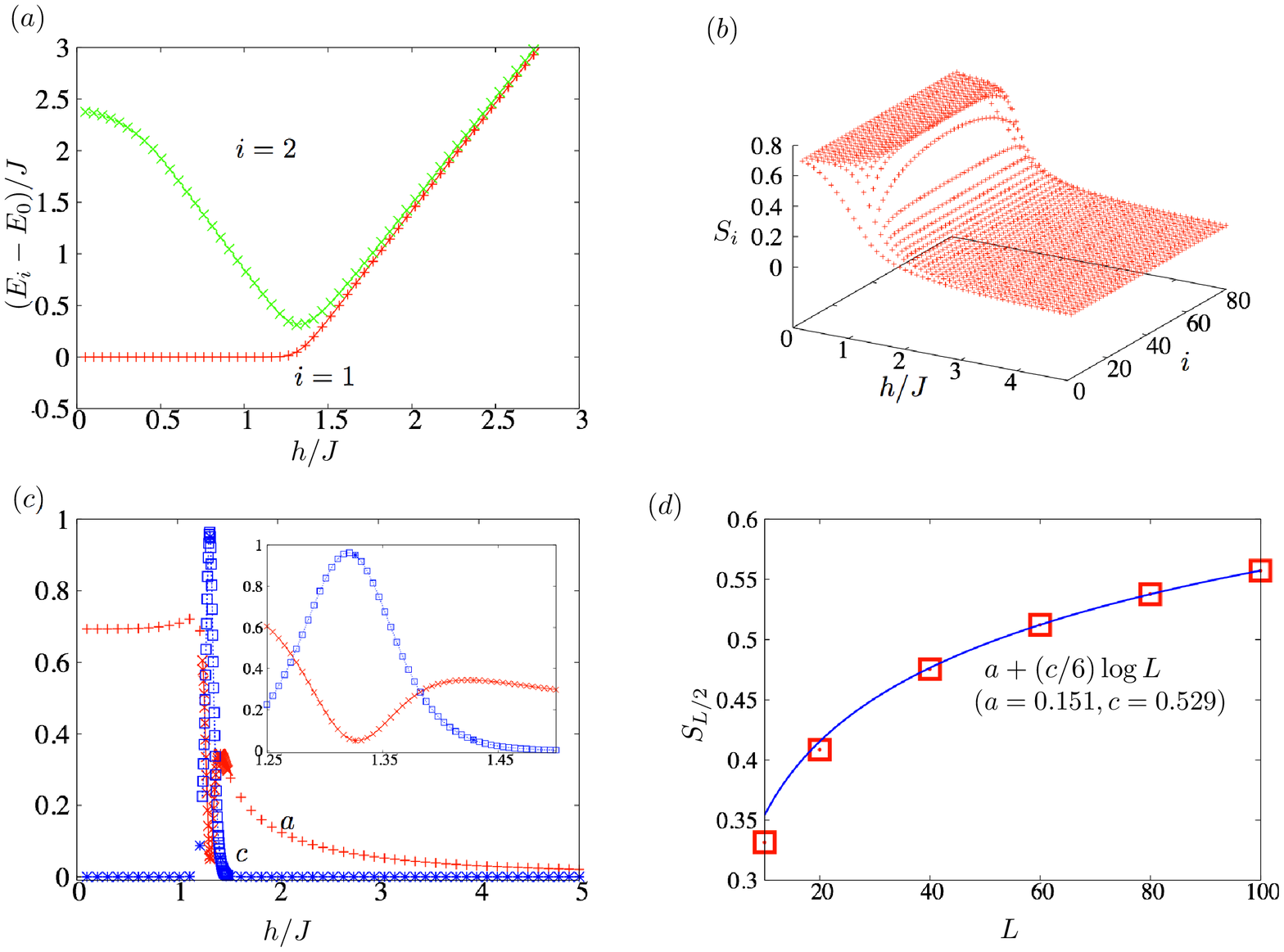}
\caption{  \label{fig:LRIsingStatics}  (color online) \emph{Statics of the dipolar Ising model.} (a) The gaps to the two-lowest-lying eigenstates, computed using eMPS.  Here we see a breakdown of the linear dispersion at small $h$, indicating interactions between quasiparticles.  (b) The bond entropy $S_i$ vs.~the site index $i$ and magnetic field $h$ for $L=80$ sites.  The point of greatest curvature is shifted towards larger $h$ with respect to the Ising model.  (c) The central charge $c$ (blue line) and bulk entanglement $a$ (red line) extracted from a fit to Eq.~(\ref{eq:finiteCC}) vs.~$h$ for $L=80$ sites, neglecting 30 sites at both boundaries.  The inset is a close-up of the critical region.  (d) The bond entropy of the central site $S_{L/2}$ vs.~$L$ at $h=1.362$ together with a fit extracting the central charge.}
\end{center}
\end{figure}

We first turn to the statics of this model, shown in Fig.~\ref{fig:LRIsingStatics}. Many of the features are similar to those of the nearest-neighbor Ising model.  The most important differences are that the critical region is shifted towards larger $h$ with respect to the nearest-neighbor Ising model as seen in panels b) and c).  This indicates increased stability of the ferromagnetic phase, in accordance with expectations.  Using these points as a guide, we determine  the critical point to be $h_c=1.362\pm0.01$, as shown by the scaling in panel d).  Additionally, the energy of the first even parity excited state deviates from pure $z=1$ linear behavior near $h=0$, indicating interactions between quasiparticles which were noninteracting in the nearest-neighbor Ising model.

\begin{figure}[t]
\begin{center}
\epsfxsize=0.8\textwidth\leavevmode \epsfbox{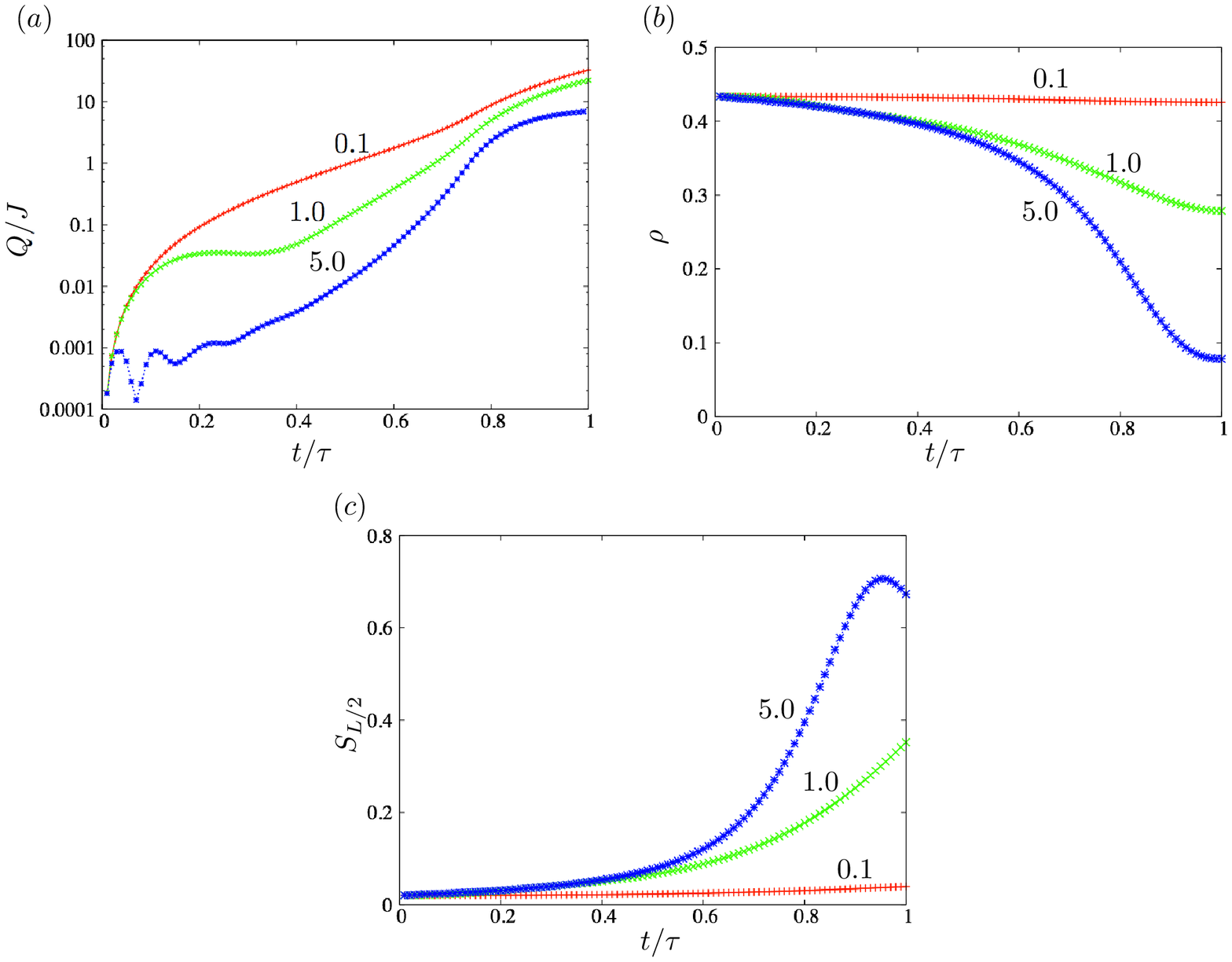}
\caption{  \label{fig:LRIsingDynamics}  (color online) \emph{Dynamics of the dipolar Ising model.} (a) The heat as a function of time displays a slower buildup in post-critical non-adiabatic effects for longer quenches, as in the nearest-neighbor case. (b) The density of domain walls as a function of time is comparable to that for the nearest-neighbor case, but no longer has the same interpretation in terms of quasiparticles.  (c) The bond entropy of the central bond oscillates for the shorter quench time $J\tau/\hbar=5$ due to the quench passing the critical point sooner than in the nearest-neighbor case.}
\end{center}
\end{figure}

We now turn to dynamics, following the same quenching protocol Eq.~(\ref{eq:Quenchprotocol}) as above.  The results are shown in Fig.~\ref{fig:LRIsingDynamics}.  We reiterate that the dynamics of this model cannot be handled by Bogoliubov-de Gennes methods, nor straightforwardly with standard tDMRG/TEBD.  The density of defects no longer represents the density of quasiparticles at the critical point, but we compute it for comparison with the results of the nearest-neighbor Ising model.  Because of the larger MPO bond dimensions and the more rapid growth of bond dimension for this model, we restrict our analysis to short times $J\tau/\hbar \le 5$, though an optimized implementation could reach longer times.  The basic features are similar to the dynamics of the nearest-neighbor Ising model.  One quantitative difference is that, because the critical point is reached at an earlier time than in the nearest-neighbor Ising model, oscillations in the central bond entropy occur for shorter quench times.

\section{Conclusions}
\label{sec:concl}
The power of matrix product state algorithms over DMRG-based algorithms is most readily apparent when multiple states are involved, as each state may be represented as a separate matrix product state in the former approach.  Because matrix product states with a fixed bond dimension do not form a vector space, a set of matrix product states carries more information at smaller numerical cost than the same set represented as a multi-state targeted basis in DMRG.  We have presented two algorithms, eMPS and tMPS, which use this property to find eigenstates and perform time evolution of strongly correlated 1D quantum systems. 

eMPS uses a set of eigenstates stored as separate matrix product states to define a projector into the space orthogonal to this set.  We use this projector to explicitly orthogonalize a variational state against previously determined eigenvectors in order to find excited states.  The explicit matrix product state representation allows us to store the excited states much more accurately than with standard DMRG, and allows allows us to ensure global orthogonality between the eigenstates to a desired tolerance.  The variance, which is computed exactly using the matrix product operator representation of the Hamiltonian, gives strict error bars on the energies obtained with this procedure.

tMPS avoids the need for an explicit representation of the propagator by using a commutator-free Magnus expansion and then building successive Krylov subspace approximations to the matrix exponentials which appear in the expansion.  Each vector in the Krylov subspace is stored as a separate matrix product state to maximize efficiency.  Furthermore, the operators $\hat{\Omega}_i$ have exact representations as matrix product operators with the same bond dimension as the Hamiltonian.  Our algorithm eliminates the need for Hamiltonian-specialized implementation of dynamics.  Additionally, by carefully accounting for the time dependence of the Hamiltonian with a commutator-free Magnus expansion, the error in our algorithm depends only on commutators of the Hamiltonian with itself at different times and not on its derivatives.  As with eMPS, the errors are rigorously accounted for by considering distance functionals with the variational state.

The matrix product operator forms of 1D Hamiltonians can be obtained using a small set of finite state automaton rules such as the site, bond, and finite function rules.  Using matrix product operator arithmetic, we can add together the various terms in a Hamiltonian from these rules to form a complete canonical MPO representation.  This representation allows for templating of Hamiltonians which depend only on the type of interactions and not on the microscopic constituents of the lattice model.  Furthermore, given the time-dependent form of the Hamiltonian parameters, one can use the same template to form the operators $\hat{\Omega}_i$ which appear in tMPS at negligible numerical cost.

We used our algorithms to study both the nearest-neighbor Ising model in a transverse field and a long-range Ising model in a transverse field.  By the closing of the gap obtained with eMPS we determined that the critical points of these models were conformal, and so we used the Calabrese-Cardy formula for the bond entropy of conformal systems to locate the critical point and its associated central charge.  The known result $h_c=1$ was verified for the nearest-neighbor case, and the critical point was shifted deeper into the paramagnetic region $h_c=1.362\pm0.01$ for the long-range case, indicating a stabilization of the ferromagnetic phase.  We used tMPS to study the dynamics of these models following a linear quench of the transverse field from the paramagnetic phase through the critical point into the ferromagnetic phase.  We saw strong non-adiabatic effects in the heat as quenching continued into the ferromagnetic region, scaling of the density of defects consistent with the Kibble-Zurek hypothesis, and the oscillation of the bond entropy near its limiting ferromagnetic phase value for nearly adiabatic quenches.

This work was supported by the Heidelberg Center for Quantum Dynamics, the Alexander von Humboldt Foundation and the National Science Foundation under Grant PHY-0903457.  We also acknowledge the Golden Energy Computing Organization at the Colorado School of Mines for the use of resources acquired with financial assistance from the National Science Foundation and the National Renewable Energy Laboratories.  We thank Erman Bekaroglu for discussions and a thorough reading of the manuscript.

\newpage
\appendix
\section{Time evolution of exact solution of 1D transverse-field quantum Ising model for comparison with tMPS}
\label{sec:IsingExact}

The solution of the statics of the transverse-field quantum Ising model is covered in standard texts~\cite{sachdev1999}. However, for comparison with tMPS we require a description of the dynamics, and so we present the dynamical case here.  To find the exact solution of the Ising model,
\begin{eqnarray}
\hat{H}=-J\sum_{\langle i,j\rangle}\hat{\sigma}_i^{z}\hat{\sigma}_{j}^z-h\sum_{i=1}^{L}\hat{\sigma}^x_i\, ,
\end{eqnarray}
we affect the Jordan-Wigner transformation
\begin{eqnarray}
\hat{\sigma}_i^x=1-2\hat{c}_i^{\dagger}\hat{c}_i\, ,\\
\hat{\sigma}_i^z=-\left(\hat{c}_i+\hat{c}_i^{\dagger}\right)\prod_{j<i}\left(1-2\hat{c}_j^{\dagger}\hat{c}_j\right)\, ,
\end{eqnarray}
where the fermionic operators $\hat{c}_i$ satisfy the anticommutation relations $\{\hat{c}_i,\hat{c}_j^{\dagger}\}=\delta_{ij}$, $\{\hat{c}_i,\hat{c}_j\}=\{\hat{c}_i^{\dagger},\hat{c}_j^{\dagger}\}=0$.  This transforms the Ising model into the fermion Hamiltonian
\begin{eqnarray}
\label{eq:fermiH}\hat{H}=-J\sum_{i=1}^{L-1}\left(\hat{c}_i^{\dagger}\hat{c}_{i+1}^{\dagger}+\hat{c}_i^{\dagger}\hat{c}_{i+1}+\mathrm{h.c.}\right)+2h\sum_{i=1}^{L}\hat{c}_i^{\dagger}\hat{c}_i-Lh\, .
\end{eqnarray}
As this is a quadratic form in fermion operators, it may be diagonalized by a canonical (Bogoliubov) transformation~\cite{Lieb_Schultz_61,Pfeuty_70}
\begin{eqnarray}
\hat{c}_i=\sum_{k=1}^{L}\left(u_{ik}\hat{\gamma}_k+v_{ik}^{\star}\hat{\gamma}_k^{\dagger}\right)
\end{eqnarray}
which provides the set of Bogoliubov-de Gennes equations
\begin{eqnarray}
\epsilon_k \mathbf{u}_k=A\mathbf{u}_k+B\mathbf{v}_k\\
\epsilon_k \mathbf{v}_k=-B\mathbf{u}_k-A\mathbf{v}_k
\end{eqnarray}
where $\mathbf{u}_k$ are the elements of $\{u_{ik},i=1,\dots,L\}$ arranged as a vector and likewise for $\mathbf{v}_k$.  The matrices $A$ and $B$ are real and tridiagonal, with the nonzero matrix elements $A_{i,i}=2h$, $A_{i,j}=-J$, $|i-j|=1$ and $B_{i,i+1}=-B_{i+1,i}=-J$. The pairs $(u_{ik},v_{ik})$ with positive energy $\epsilon_k$, $\epsilon_1\le \epsilon_2\le\dots \le \epsilon_L$, and the normalization $|\mathbf{u}_k|^2+|\mathbf{v}_k|^2=1$ define quasiparticle operators
\begin{eqnarray}
\hat{\gamma}_k=u_{ik}^{\star}\hat{c}_i+v_{ik}\hat{c}^{\dagger}_i
\end{eqnarray}
which bring the Hamiltonian into the diagonal form
\begin{eqnarray}
\hat{H}=\sum_{k=1}^{N}\epsilon_k\left(\hat{\gamma}_k^{\dagger}\hat{\gamma}_k-\frac{1}{2}\right)\, .
\end{eqnarray}
Corresponding to every such pair is another pair $(\tilde{u}_{ik},\tilde{v}_{ik})=(v_{ik},u_{ik})$ with $\tilde{\epsilon}_k=-\epsilon_k$ which defines the conjugate operator $\hat{\gamma}_k^{\dagger}$.  Writing $\mathbf{u}_k$ and $\mathbf{v}_k$ together as a composite vector, the Bogoliubov-de Gennes equations take the form of a real symmetric eigenproblem of dimension $2L$:
\begin{eqnarray}
\epsilon_{k}\left(\begin{array}{c} \mathbf{u}_k\\ \mathbf{v}_k\end{array}\right)=\left(\begin{array}{cc} A&B\\ -B&-A\end{array}\right)\left(\begin{array}{c} \mathbf{u}_k\\ \mathbf{v}_k\end{array}\right)
\end{eqnarray}
which can be readily solved using standard eigenvalue methods~\cite{lapack}.

Evolution under the fermion Hamiltonian Eq.~(\ref{eq:fermiH}) does not preserve the number of fermions $N_F$ but it does preserve the fermionic parity $(-1)^{N_F}$.  Because the ground state is the Bogoliubov vacuum it contains no fermions, and so the first \emph{accessible} excited state consists of two Bogoliubov excitations, one in each of the lowest two modes.  The gap between the ground and first excited states is thus $\epsilon_1+\epsilon_2$.

\begin{figure}[t]
\begin{center}
\epsfxsize=0.4\textwidth \leavevmode \epsfbox{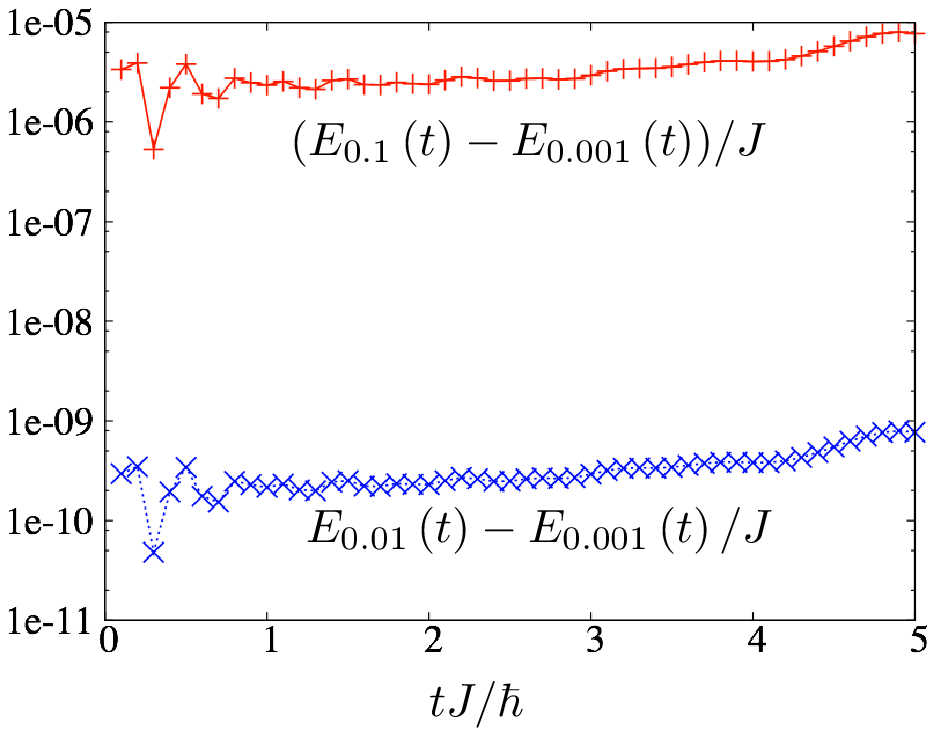}
\caption{  \label{fig:NumExact}  \emph{Demonstration of Bogoliubov-de Gennes method convergence.} The errors in the energy computed with a given time step $\delta t$, $E_{\delta t}$, are shown as a function of time.  The ratio of the errors as a function of time is roughly $(0.01/0.1)^4=0.0001$, as should be expected for our fourth-order CFME.  Errors in the density of defects behave similarly.}
\end{center}
\end{figure}

We now consider the Heisenberg equations of motion for the fermi operators
\begin{equation}
i\hbar \frac{d\hat{\mathbf{c}}\left(t\right)}{dt}=A\left(t\right)\hat{\mathbf{c}}\left(t\right)+B\left(t\right)\hat{\mathbf{c}}^{\dagger}\left(t\right)\, ,
\end{equation}
where $A(t)$ and $B(t)$ are the time-dependent generalizations of $A$ and $B$ above and $\hat{\mathbf{c}}(t)$ and $\hat{\mathbf{c}}^{\dagger}(t)$ are the elements of $\{\hat{c}_i(t)\}$ and $\{\hat{c}_i^{\dagger}(t)\}$, respectively, arranged as vectors.  Because this equation is linear in the Fermi operators it may be diagonalized with a time-dependent Bogoliubov transformation
\begin{equation}
\hat{c}_i\left(t\right)=\sum_{k=1}^{L}\left(u_{ik}\left(t\right)\gamma_k+v_{ik}^{\star}\left(t\right)\hat{\gamma}_k^{\dagger}\right)
\end{equation}
where $\mathbf{u}_i(t)$ and $\mathbf{v}_i(t)$ subject to the time-dependent Bogoliubov-de Gennes equations
\begin{eqnarray}
\label{eq:TDBDG}i\hbar\frac{d}{dt}\left(\begin{array}{c} \mathbf{u}_k\left(t\right)\\ \mathbf{v}_k\left(t\right)\end{array}\right)=\left(\begin{array}{cc} A\left(t\right)&B\left(t\right)\\ -B\left(t\right)&-A\left(t\right)\end{array}\right)\left(\begin{array}{c} \mathbf{u}_k\left(t\right)\\ \mathbf{v}_k\left(t\right)\end{array}\right)\equiv \mathcal{H}\left(t\right)\left(\begin{array}{c} \mathbf{u}_k\left(t\right)\\ \mathbf{v}_k\left(t\right)\end{array}\right)\, ,
\end{eqnarray}
and $\hat{\gamma}_k$ and $\hat{\gamma}_k^{\dagger}$ diagonalize the Hamiltonian at the initial time.  Equivalently, $u$ and $v$ define time-dependent Bogoliubov operators
\begin{eqnarray}
\hat{\gamma}_k\left(t\right)=u^{\star}_{ik}\left(t\right)\hat{c}_i+v_{ik}\left(t\right)\hat{c}_i^{\dagger}
\end{eqnarray}
such that the time-evolved state $|\psi(t)\rangle$ is the Bogoliubov vacuum of this set, i.e.~$\hat{\gamma}_k(t)|\psi(t)\rangle=0$.
To compare with the MPS simulations, we note that the energy at time $t$ is
\begin{eqnarray}
-\frac{1}{2}\sum_k\epsilon_k\left(t\right)=-\frac{1}{2}\sum_k \left( \mathbf{u}_k\left(t\right)\mathbf{v}_k\left(t\right)\right)\mathcal{H}\left(t\right)\left(\begin{array}{c} \mathbf{u}_k\left(t\right)\\ \mathbf{v}_k\left(t\right)\end{array}\right)\, .
\end{eqnarray}
Similarly, the density of defects is
\begin{eqnarray}
\rho\left(t\right)=\frac{1}{2L}\sum_{i=1}^{L-1}\langle \psi\left(t\right)|\left(1-\hat{\sigma}^z_i\hat{\sigma}^z_{i+1}\right)|\psi\left(t\right)\rangle\, ,\\
=\frac{1}{2L}\sum_{i=1}^{L-1}\left[1-\sum_{k=1}^{L}\left(v_{i,k}\left(t\right)-u_{i,k}\left(t\right)\right)\left(u_{i+1,k}^{\star}\left(t\right)+v_{i+1,k}^{\star}\left(t\right)\right)\right]\, .
\end{eqnarray}

Time evolution thus reduces to the solution of a $2L\times2L$ time-dependent matrix differential equation which we solve using a CFME as in Sec.~\ref{sec:CFME}.  Because the dimensions of the system are much smaller than those of the associated MPS problem we are able to take very small time steps, and so the results obtained in the method may be considered to be numerically exact, see Fig.\ref{fig:NumExact}.  Here we demonstrate that the error incurred in the energy as a function of time scales with the infinitesimal time step $\delta t$ as $\delta t^4$ using our fourth-order CMFE.  Hence, by decreasing $\delta t$, any desired degree of accuracy may be met.

\newpage

\bibliographystyle{prsty}
\bibliography{running}

\end{document}